\documentclass[fleqn,twoside]{revtex4}

\usepackage{amsmath}
\usepackage{graphicx}
\usepackage{psfrag}
\usepackage{fancyhdr}
\usepackage{amsfonts}
\usepackage{amssymb}

\DeclareMathOperator{\tr}{Tr}

\def\slash#1{\setbox0=\hbox{$#1$}               % set a box for #1
        \dimen0=\wd0                            % and get its size
        \setbox1=\hbox{/} \dimen1=\wd1          % get size of /
        \ifdim\dimen0>\dimen1                   % #1 is bigger
        \rlap{\hbox to \dimen0{\hfil/\hfil}}    % so center / in box
        #1                                      % and print #1
        \else                                   % / is bigger
        \rlap{\hbox to \dimen1{\hfil$#1$\hfil}} % so center #1
        /                                       % and print /
        \fi}                                    %

\newcommand{\be}{\begin{equation}} \newcommand{\ee}{\end{equation}}
\newcommand{\ba}{\begin{eqnarray}} \newcommand{\ea}{\end{eqnarray}}
\newcommand{\bea}{\begin{eqnarray}} \newcommand{\eea}{\end{eqnarray}}
\newcommand{\bean}{\begin{eqnarray*}} \newcommand{\eean}{\end{eqnarray*}}
\newcommand{\st}{{\scriptscriptstyle T}}
\newcommand{\sT}{{\scriptscriptstyle T}}

\newcommand{\sN}{{\scriptscriptstyle N}}
\newcommand{\sS}{{\scriptscriptstyle S}}
\newcommand{\sL}{{\scriptscriptstyle L}}

\begin{document}

\title{Single spin asymmetries in hadron-hadron collisions}

\author{A. Bacchetta$^1$}
\email{alessandro.bacchetta@physik.uni-regensburg.de}
\affiliation{
$^1$Institut f\"ur Theoretische Physik, Universit\"at Regensburg,\\ 
D-93040 Regensburg, Germany
\\and\\
$^2$Department of Physics and Astronomy, Vrije Universiteit Amsterdam,\\
NL-1081 HV Amsterdam, the Netherlands
}

\author{C.J. Bomhof$^2$}
\email{cbomhof@nat.vu.nl}
\affiliation{
$^1$Institut f\"ur Theoretische Physik, Universit\"at Regensburg,\\ 
D-93040 Regensburg, Germany
\\and\\
$^2$Department of Physics and Astronomy, Vrije Universiteit Amsterdam,\\
NL-1081 HV Amsterdam, the Netherlands
}

\author{P.J. Mulders$^2$}
\email{pjg.mulders@few.vu.nl}
\affiliation{
$^1$Institut f\"ur Theoretische Physik, Universit\"at Regensburg,\\ 
D-93040 Regensburg, Germany
\\and\\
$^2$Department of Physics and Astronomy, Vrije Universiteit Amsterdam,\\
NL-1081 HV Amsterdam, the Netherlands
}
                                                                                
\author{F. Pijlman$^2$}
\email{f.pijlman@few.vu.nl}
\affiliation{
$^1$Institut f\"ur Theoretische Physik, Universit\"at Regensburg,\\ 
D-93040 Regensburg, Germany
\\and\\
$^2$Department of Physics and Astronomy, Vrije Universiteit Amsterdam,\\
NL-1081 HV Amsterdam, the Netherlands
}

%\pacs{13.60.Hb,13.88.+e,12.39.Fe}

\begin{abstract}
We study weighted azimuthal single spin asymmetries in hadron-hadron 
scattering using the diagrammatic approach at leading order and assuming
factorization. The effects of the intrinsic 
transverse momenta of the partons are taken into account.
We show that the way in which $T$-odd functions, such as
the Sivers function, appear in these processes does not merely involve 
a sign flip when compared with semi-inclusive deep inelastic 
scattering, such as in the case of the Drell-Yan process.
Expressions for the weighted scattering cross sections in terms of
distribution and fragmentation functions folded with hard cross sections
are obtained by introducing modified hard cross sections, referred to as 
gluonic pole cross sections.
\end{abstract}
\date{\today}
\maketitle

\section{Introduction}

Accessing the effects arising from the transverse momentum of quarks in
hadrons requires hard processes involving, at least, two hadrons (or hadronic
jets) and a hard scale to separate them. 
This is most cleanly achieved in electroweak
processes in which the gauge boson provides the hard scale separating
the two hadronic regions. The transition between the hadronic regions 
and the hard subprocess is described by soft quark and gluon correlation
functions, implying approximate collinearity between the quarks, 
gluons and hadrons involved. 
Without effects of quark intrinsic transverse momentum these are bilocal, 
lightlike separated, 
matrix elements where collinear gluons provide the gauge-link. 
Transverse momentum dependent correlation
functions involve bilocal matrix elements off the 
lightcone~\cite{Ralston:1979ys}. Here the issue of color gauge-invariance
is slightly more complex, involving gauge fields at lightcone 
infinity~\cite{Boer:1999si,Belitsky:2002sm,Collins:2002kn, 
Collins:2003fm, Boer:2003cm}.
The gauge-link structure may depend on the hard subprocess and 
leads to observable consequences.

Absorbing the soft physics in the correlation functions requires 
coupling of, essentially, collinear quarks (and gluons) to the hadronic
region. These partons themselves are approximately on mass-shell. 
In the absence of a hard scale from an electroweak boson, as 
in strong interaction processes, the simplest hard subprocess (large
momentum transfer) involving on-shell quarks and gluons is a two-to-two process.
 
In this paper we discuss hard hadron-hadron scattering
processes using the diagrammatic approach rather than the commonly used
helicity approach~\cite{Jaffe:1996ik,Anselmino:1999pw,Anselmino:2002pd,
Boer:2003tx,D'Alesio:2004up,Anselmino:2004ky}.
This has the advantage that we can directly connect to
the matrix elements of quark and gluon fields, without having to go 
through the step of rewriting them into parton distributions with 
specific helicities. It allows us to include the effects of collinear 
gluons, determining the gauge-link structure and to compare this for
semi-inclusive deep inelastic scattering (SIDIS) and the Drell-Yan
process (DY). We note that in this paper we assume the validity 
of factorization.

We will consider the possibilities to measure transverse moments
which are obtained from transverse momentum dependent (TMD) distribution and
fragmentation functions upon integration over intrinsic transverse momentum 
($k_\st$) including a $k_\st$-weighting. 
In the transverse moments the effects of the gauge-link structure 
remain visible. 
For this one needs to classify the distribution and fragmentation functions 
as $T$-even or $T$-odd. In single-spin
asymmetries at least one (in general an odd number of) $T$-odd
function appears, while in unpolarized processes or double-spin
asymmetries an even number of $T$-odd functions must appear. 
The importance of considering transverse momentum dependence
comes from the fact that for spin 0 and spin $\frac{1}{2}$ hadrons the 
simple transverse momentum integrated distribution and fragmentation functions,
relevant at leading order, are all $T$-even.

The specific hadronic process that we will consider is the 
2-particle inclusive process $H_1{+}H_2\rightarrow h_1{+}h_2{+}X$, which in 
order to separate the hadronic regions requires minimally a two-to-two 
hard subprocess. Also included are inclusive hadron-jet and jet-jet 
production in hadron-hadron scattering.
The 1-particle inclusive process $p^\uparrow{+}p\rightarrow\pi{+}X$ involving
a transversely polarized proton is known to show a large single-spin 
asymmetry~\cite{Adams:1991rw,Adams:1991cs,Bravar:1996ki,
Adams:2003fx,Adler:2003pb}. Some of the 
mechanisms~\cite{Sivers:1989cc,Sivers:1990fh,Qiu:1991wg,Collins:1992kk,Kanazawa:2000hz} 
to explain these asymmetries involve $T$-odd functions, such as the Sivers
distribution function or the Collins fragmentation 
function~\cite{Boglione:1999pz}. These functions 
are expected to appear in a cleaner way in 2-particle inclusive 
processes~\cite{Boer:2003tx}.
Here we only consider single-hadron fragmentation functions,
in which case the 2-particle inclusive production requires $h_1$ and
$h_2$ to belong to different, in the perpendicular plane approximately 
opposite, jets. 

In this paper we limit ourselves to the (anti)quark contributions with as main
goal to show the relevant gauge-link structure for the $T$-odd Sivers
distribution functions $f_{1T}^{(1)}$ and the Collins fragmentation functions
$H_1^{\perp (1)}$ entering these processes. This is important 
for the study of universality of these functions.
The paper is structured as follows. 
In section~\ref{Kinematics} we consider the kinematics 
particular to $2{\rightarrow}2$ particle scattering. 
In section~\ref{CrossSec} we discuss our approach and several (weighted)
scattering cross section are written down for hadronic pion production and
hadronic jet production in section~\ref{SSA}.
Details about the gauge-links and their consequences for distribution and 
fragmentation functions are dealt with in the appendices.

\section{kinematics\label{Kinematics}}

\begin{figure}
\includegraphics{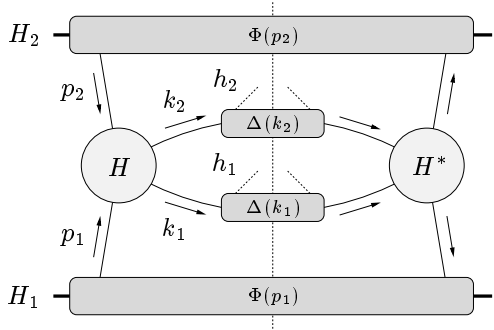}
\caption{The leading order contribution to the cross section of $H_1{+}H_2\rightarrow h_1{+}h_2{+}X$.\label{PPpipi}}
\end{figure}

The hard scale in the process
$H_1(P_1){+}H_2(P_2)\rightarrow h_1(K_1){+}h_2(K_2){+}X$
is set by the center-of-mass energy $\sqrt s=E^\text{cm}$.
The leading order contribution to the scattering cross section is shown in Fig.~\ref{PPpipi}. 
In a hard scattering process it is 
important to get as much information about the partonic momenta as possible, 
in our case including, in particular, their transverse momenta. 
The partonic momenta, for which $p_1{\cdot}P_1\sim p_1^2\sim P_1^2=M_1^2$ 
are of hadronic scale, are expanded as follows
\begin{subequations}\label{PartonMomenta}
\begin{align}
p_1&=x_1\,P_1+\sigma_1\,n_1+p_{1\st} \ ,\label{PartonMomentaA}\\
p_2&=x_2\,P_2+\sigma_2\,n_2+p_{2\st}\ ,\\
k_1&=z_1^{-1}\,K_1+\sigma_1'\,n_1'+k_{1\st}\ ,\\
k_2&=z_2^{-1}\,K_2+\sigma_2'\,n_2'+k_{2\st}\ ,
\end{align}
\end{subequations}
where the $n_i$ ($n_i'$) are lightlike vectors chosen such that 
$P_1\cdot n_1\propto\mathcal O(s^{1/2})$ 
and similarly for the other partonic momenta. 
The fractions $x_i = p_i{\cdot}n_i/P_i{\cdot}n_i$ and 
$z_i^{-1}= k_i{\cdot}n_i'/K_i{\cdot}n_i'$ are lightcone momentum fractions.
The quantities multiplying the vectors $n_i$ are of order $s^{-1/2}$ and are 
the lightcone components conjugate to $p_i{\cdot}n_i$. 
They are given by
\begin{equation}
\sigma_i=\frac{p_i{\cdot}P_i-x_i\,M_i^2}{P_i{\cdot}n_i}\ ,
\end{equation}
with similar expressions for the $\sigma_i'$.
If any of the `parton' momenta is actually an external momentum 
(for leptons or when describing jets) 
the momentum fractions become unity and the transverse momenta 
and $\sigma_i$ vanish.

Integration over parton momenta is written as
\be
d^4p_1 = dx_1\,d^2p_{1\st}\,d(p_1{\cdot}P_1)\ ,
\ee
with $d(p_1{\cdot}P_1)=(P_1{\cdot}n_1)\,d\sigma_1$ 
and similar expressions for $d^4p_2$, $d^4k_1$ and $d^4k_2$.
The integrations over the parton momentum components $(p_i{\cdot}P_i)$ 
and $(k_i{\cdot}K_i)$ will be included in the definitions
of the TMD distribution and fragmentation functions.
Note that the subscripts $T$ have a different meaning in each of the 
above decompositions, i.e., $p_{1\st}$ is transverse to $P_1$ and $n_1$,
while $p_{2\st}$ is transverse to $P_2$ and $n_2$, etc.
Momentum conservation relates the partonic momenta:
\begin{equation}\label{momconservation}
p_1+p_2-k_1-k_2=0\ . 
\end{equation}
This relation, however, 
does not imply that the sum of the intrinsic transverse momenta,
$q_{\st}\equiv p_{1\st}{+}p_{2\st}{-}k_{1\st}{-}k_{2\st}$ 
$\approx z_1^{-1}K_1{+}z_2^{-1}K_2{-}x_1P_1{-}x_2P_2$
vanishes.

\begin{figure}
\centering
\includegraphics[width=5cm]{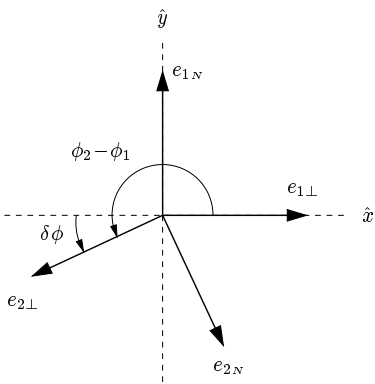}
\caption{\label{PERPplane}
Plane perpendicular to the incoming hadronic momenta.}
\end{figure}

We use the incoming momenta $P_1$ and $P_2$ to define perpendicular
momenta $K_{i\perp}$ orthogonal to the incoming hadronic momenta,
$K_{i\perp}{\cdot}P_1=K_{i\perp}{\cdot}P_2=0$. For the perpendicular
momenta it is convenient to scale the variables using
$x_{i\perp}=2\vert K_{i\perp}\vert/\sqrt s$. 
For the outgoing hadrons we use the pseudo-rapidities 
$\eta_i$ defined by $\eta_i=-\ln\tan(\frac{1}{2}\theta_i)$ where the 
$\theta_i$ are the polar angles of these hadrons in the center-of-mass frame.
All invariants involving the external momenta can be expressed in terms of 
these variables:
\begin{subequations}
\begin{alignat}{2}
P_1\cdot K_1&=\tfrac{1}{4}s\,x_{1\perp}\,e^{-\eta_1}\ ,&\qquad
P_2\cdot K_1&=\tfrac{1}{4}s\,x_{1\perp}\,e^{+\eta_1}\ ,\\
P_1\cdot K_2&=\tfrac{1}{4}s\,x_{2\perp}\,e^{-\eta_2}\ ,&
P_2\cdot K_2&=\tfrac{1}{4}s\,x_{2\perp}\,e^{+\eta_2}\ ,
\end{alignat}
and $P_1{\cdot}P_2=\frac{1}{2}s$.
These identities are valid up to subleading order in $\sqrt s$.
The remaining invariant $K_1{\cdot}K_2$ is not independent of the others. 
\emph{To leading order}, one has
\begin{equation}\label{Sprime}
K_1\cdot K_2
=\tfrac{1}{2}s\,x_{1\perp}x_{2\perp}\,
\cosh^2\big[\tfrac{1}{2}(\eta_1{-}\eta_2)\big]\ .
\end{equation}
\end{subequations}
To subleading order the outgoing hadronic momenta can now be written as
\begin{subequations}
\begin{align}
K_1
&=\frac{(K_1{\cdot}P_2)\,P_1 + (K_1{\cdot}P_1)\,P_2}{P_1{\cdot}P_2}+K_{1\perp}
=\tfrac{1}{2}x_{1\perp}\,
\big(\,e^{+\eta_1}P_1+e^{-\eta_1}P_2\,\big)+K_{1\perp}\ ,\\
K_2
&=\frac{(K_2{\cdot}P_2)\,P_1 + (K_2{\cdot}P_1)\,P_2}{P_1{\cdot}P_2}+K_{2\perp}
=\tfrac{1}{2}x_{2\perp}\,
\big(\,e^{+\eta_2}P_1+e^{-\eta_2}P_2\,\big)+K_{2\perp}\ .
\end{align}
\end{subequations}
The two perpendicular vectors $K_{1\perp}$ and $K_{2\perp}$ 
are approximately back-to-back (see Fig.~\ref{PERPplane}).
Sometimes the Feynman variables
$x_{iF}=K_{iz}^\text{cm}/K_{iz}^\text{cm(max)}=x_{i\perp}\sinh\eta_i$ 
are used in the literature.
Another useful variable in writing down cross sections is the quantity $y$, 
which is defined via the Mandelstam variables of the partonic subprocess, 
and can be related to the pseudo-rapidities,
\begin{equation}\label{Y}
y=-\frac{\hat t}{\hat s}
=\sqrt{\frac{(P_1{\cdot}K_1)(P_2{\cdot}K_2)}{(P_1{\cdot}P_2)(K_1{\cdot}K_2)}}
=\frac{1}{e^{(\eta_1-\eta_2)}+1}\ .
\end{equation}

Dividing $K_{1\perp}$ and $K_{2\perp}$ by the momentum fractions one 
immediately sees from the decompositions of the partonic momenta that the 
vector 
\begin{equation}
r_\perp=\frac{K_{1\perp}}{z_1}+\frac{K_{2\perp}}{z_2}\ ,
\end{equation}
only involves transverse momenta of partons. 
It is just the small projection of the transverse momentum in the 
perpendicular plane, 
$r_\perp \approx q_{\st\perp}$. 
The vectors $K_{i\perp}$ themselves are not `small' vectors. They are 
spacelike vectors with invariant length of $\mathcal O(\sqrt s)$.
In the analysis of the kinematics in the transverse plane the momentum
fractions are not direct observables. 
In particular, experimentally it is more convenient to work with the 
directions of the vectors $K_{i\perp}$ and the corresponding orthogonal 
directions (see Fig.~\ref{PERPplane})
\begin{equation}
e_{1\perp}^\mu=\frac{K_{1\perp}^\mu}{\vert K_{1\perp}\vert}\ ,
\qquad
e_{1\sN}^\mu
=-\frac{2}{s}\,\frac{\epsilon^{P_1P_2K_1\,\mu}}{\vert K_{1\perp}\vert}
=\epsilon_\perp^{\mu\nu}\,e_{1\perp\,\nu}\ ,
\end{equation}
and similarly for $K_{2\perp}$. 
As illustrated in Fig.~\ref{PERPplane}, the direction $e_{2\perp}$ is, 
up to a (small)angle $\delta\phi\equiv\phi_2{-}\phi_1{-}\pi
\propto\mathcal O(\vert p_\st\vert/\sqrt{s})$, opposite to $e_{1\perp}$.

The momentum conservation relation in Eq.~\eqref{momconservation} is enforced 
by a delta function in the scattering cross section.
The delta function can be decomposed using the basis constructed in the 
previous paragraph.
For $R=p_1{+}p_2{-}k_1{-}k_2$ this decomposition reads
\begin{equation}
\delta^4(R)
=\tfrac{1}{2}s\,\delta(R{\cdot}P_1)\,\delta(R{\cdot}P_2)\,
\delta^2(R_\perp)
=\tfrac{1}{2}s\,\delta(R{\cdot}P_1)\,\delta(R{\cdot}P_2)\,
\delta^2(q_{\st\perp}{-}r_\perp)\ .
\label{deltafunction}
\end{equation}
The arguments of the first two delta functions involve large momenta and can 
be used to relate the momentum fractions $x_1$ and $x_2$ to kinematical 
observables. For the latter two delta functions the treatment depends
on the situation. Using the orthogonal vectors $e_{1\perp}$ and 
$e_{1\sN}$ we get, up to ${\cal O}(1/\sqrt s)$,
\begin{subequations}
\begin{align}
R{\cdot}e_{1\perp} 
&= e_{1\perp}{\cdot}q_\st 
+ \left(\frac{x_{1\perp}}{z_1}-\frac{x_{2\perp}}{z_2}\right)
\frac{\sqrt s}{2}\ ,
\\
R{\cdot}e_{1\sN} 
&= e_{1\sN}{\cdot}q_\st 
- \frac{x_{2\perp}}{z_2}\,\frac{\sqrt s}{2}\,\sin(\delta\phi)\ .
\end{align}
\end{subequations}
In the case of two-hadron or hadron-jet production ($z_2=1$),
the first delta function implies that at leading order $x_{1\perp}/z_1
\approx x_{2\perp}/z_2 \equiv x_\perp$, which is interpreted as
the scaled parton perpendicular momentum,
$x_\perp = 2\,\vert k_{2\perp}\vert/\sqrt s$. 
Using the variable $x_\perp$ as an integration variable we can write
\begin{equation}\begin{split}
\delta^4(p_1{+}p_2{-}k_1{-}k_2) 
=\frac{4}{s^2}\,\frac{1}{x_{1\perp}\,x_{2\perp}}\int dx_\perp\ 
\delta&\big(\,x_1-\tfrac{1}{2}x_\perp(e^{\eta_1}{+}e^{\eta_2})\,\big)\,
\delta\big(\,x_2-\tfrac{1}{2}x_\perp(e^{-\eta_1}{+}e^{-\eta_2})\,\big)\\ 
&\times\,
\delta\Big(\,z_1^{-1}- \frac{x_\perp}{x_{1\perp}}\,\Big)\,
\delta\Big(\,z_2^{-1}- \frac{x_\perp}{x_{2\perp}}\,\Big)\,
\delta\Big(\,\frac{e_{1\sN}{\cdot}q_\st}{\sqrt s}
-\frac{x_\perp}{2}\,\sin(\delta\phi)\,\Big)\ ,
\label{deltas}
\end{split}\end{equation}
which shows that in one- or two-particle inclusive processes we are always left
with a convolution of distribution and fragmentation functions over one 
momentum fraction or, equivalently, over the parton perpendicular momentum
variable $x_\perp$.
The last delta function shows explicitly that 
$\sin(\delta\phi) \propto 1/\sqrt s$ and that it can be used to construct
cross sections weighted with one component of the intrinsic transverse
momentum, i.e.\ $e_{1\sN}{\cdot}q_\st$.

In the case that $K_1 = k_1$ and $K_2 = k_2$ (i.e.\ $z_i=1$, $k_{i\st}=0$), 
such as in production of a lepton
pair in Drell-Yan scattering or the (idealized) production of two jets, 
the delta function $\delta(R{\cdot}e_{1\perp})$ also relates intrinsic
transverse momenta to observed momenta. Therefore, 
one can construct azimuthal asymmetries involving two components of $q_\st$. 
In fact, the product of delta functions
$\delta^2(R_\perp)=\delta(R{\cdot}e_{1\perp})\,\delta(R{\cdot}e_{1\sN})$ can be
used to weigh with the transverse momenta $p_{1\st}{+}p_{2\st}$, 
as they relate $q_\st = p_{1\st}{+}p_{2\st}$ to $q \equiv k_1{+}k_2$ 
in the orthogonal plane.
With the natural choice of the $n$-vectors in the case that only
two hadrons are involved such that $P_{1\st} = P_{2\st} = 0$, 
one obtains the familiar relation 
$p_{1\st}{+}p_{2\st} = q_\st = q{-}x_1P_1{-}x_2P_2$,
leading to
\begin{equation}\label{JJdeltafunction}
\delta^4(p_1{+}p_2{-}k_1{-}k_2) 
= \frac{2}{s}\,\delta\Big(x_1 - \frac{P_2{\cdot}q}{P_1{\cdot}P_2}\Big)
\,\delta\Big(x_2 - \frac{P_1{\cdot}q}{P_1{\cdot}P_2}\Big)
\,\delta^2\Big(p_{1\st}{+}p_{2\st}{-}q_\st(q,P_1,P_2)\Big)\ .
\end{equation}

\section{cross sections\label{CrossSec}}

The scattering cross section for $p_1p_2\rightarrow h_1h_2X$ 
(see Fig.~\ref{PPpipi}) 
at tree-level is written as 
\begin{equation}
d\sigma
=\frac{1}{2s}\,\vert \mathcal M\vert^2\,
\frac{d^3 K_1}{(2\pi)^3\,2E_{K_1}}\,
\frac{d^3 K_2}{(2\pi)^3\,2E_{K_2}}\ ,
\end{equation}
where the matrix element is expressed in terms of hard amplitudes and
correlation functions (see appendix~\ref{a}).
It is given by
\begin{equation}\begin{split}
\vert\mathcal M\vert^2
=\int\!\!dx_1d^2 p_{1\st}\,&dx_2d^2 p_{2\st}\,
dz_1^{-1}d^2 k_{1\st}\,dz_2^{-1}d^2 k_{2\st}\ 
(2\pi)^4\delta^4(p_1{+}p_2{-}k_1{-}k_2)\\
&\times\,
\tr\big\{\,\Phi(x_1, p_{1\st})\Phi(x_2, p_{2\st})
\Delta(z_1, k_{1\st})\Delta(z_2, k_{2\st})
H(p_1,p_2,k_1,k_2)H^\ast(p_1,p_2,k_1,k_2)\,\big\}\ .
\label{brackets}
\end{split}\end{equation}
The trace involves the appropriate contraction of Dirac indices in
soft and hard scattering parts.
A summation over color and quark flavors is understood.
The phase-space elements are given by
\begin{equation}
\frac{d^3K_i}{(2\pi)^3\,2E_{K_i}}
=\frac{x_{i\perp}\,s}{8\,(2\pi)^2}\,
dx_{i\perp}\,d\eta_{i}\,\frac{d\phi_{i}}{2\pi}\ .
\end{equation}
Combining the phase space integration and the delta functions coming from
partonic energy-momentum conservation, one has for back-to-back hadron-hadron
production
\begin{equation}\begin{split}
d\sigma[h_1h_2]
=\frac{1}{32\,s}\,dx_{1\perp}\,&dx_{2\perp}\,d\eta_1\,d\eta_2\,
\frac{d\phi_1}{2\pi}\,\frac{d\phi_2}{2\pi}\int dx_\perp
\int d^2 p_{1\st}\,d^2 p_{2\st}\,d^2 k_{1\st}\,d^2 k_{2\st}\ 
\delta\Big(\,\frac{e_{1\sN}{\cdot}q_\st}{\sqrt s} 
-\frac{x_\perp}{2}\,\sin(\delta\phi)\,\Big)\\ 
&\times
\tr\big\{\Phi(x_1, p_{1\st})\Phi(x_2, p_{2\st})
\Delta(z_1, k_{1\st})\Delta(z_2, k_{2\st})
\,H(p_1,p_2,k_1,k_2)H^\ast(p_1,p_2,k_1,k_2)\big\}\ .
\end{split}\end{equation}
In this expression the momentum fractions are fixed by the arguments
of the delta functions in Eq.~\eqref{deltas}, i.e.\ one has
$x_1(x_\perp,\eta_1,\eta_2)$, $x_2(x_\perp,\eta_1,\eta_2)$,
$z_1(x_\perp,x_{1\perp})$, and $z_2(x_\perp,x_{2\perp})$.
Since $x_{1\perp} \le x_\perp$ and $x_{2\perp} \le x_\perp$, the
integration over $x_\perp$ is bounded from below. 

In the hadron-jet inclusive process one has 
$\Delta(z_2,k_{2\st})=\delta(z_2{-}1)\,\delta^2(k_{2\st})\,\slash k_2
=x_\perp\,\delta(x_{2\perp}{-}x_\perp)\,\delta^2(k_{2\st})\,\slash k_2$,
which implies that $z_2 = 1$ and $x_{\perp} = x_{2\perp}$. 
The cross section becomes
\begin{equation}\begin{split}
d\sigma[h_1j_2]
=\frac{x_{2\perp}}{32\,s}\,dx_{1\perp}\,dx_{2\perp}\,d\eta_1&\,d\eta_2\,
\frac{d\phi_1}{2\pi}\,\frac{d\phi_2}{2\pi}
\int d^2 p_{1\st}\,d^2 p_{2\st}\,
d^2 k_{1\st}\ 
\delta\Big(\,\frac{e_{1\sN}{\cdot}q_\st}{\sqrt s} 
-\frac{x_{2\perp}}{2}\,\sin(\delta\phi)\,\Big)\\ 
&\times
\tr\big\{\Phi(x_1, p_{1\st})\Phi(x_2, p_{2\st})
\Delta(z_1, k_{1\st})
\,H(p_1,p_2,k_1,k_2)H^\ast(p_1,p_2,k_1,k_2)\big\}\ .
\end{split}\end{equation}
As stated in the previous section,
in back-to-back jet production both delta functions in the perpendicular plane
relate observed kinematical variables to intrinsic transverse momenta.
Therefore, in jet-jet production we use the expression in
Eq.~\eqref{JJdeltafunction} rather than Eq.~\eqref{deltas} for
the momentum conserving delta function, leading to
\begin{equation}\begin{split}
d\sigma[j_1j_2]
=\frac{x_{1\perp}\,x_{2\perp}}{64}\,
dx_{1\perp}\,dx_{2\perp}\,d\eta_1\,d\eta_2\,
&\frac{d\phi_1}{2\pi}\,\frac{d\phi_2}{2\pi}
\int d^2 p_{1\st}\,d^2 p_{2\st}\
\delta^2(p_{1\st}{+}p_{2\st}{-}q_{\st})\\
&\times
\tr\big\{\Phi(x_1, p_{1\st})\Phi(x_2, p_{2\st})
\,H(p_1,p_2,k_1,k_2)H^\ast(p_1,p_2,k_1,k_2)\big\}\ ,
\end{split}\end{equation}
where $z_1 = z_2 = 1$, $x_{1\perp} = x_{2\perp} = x_\perp$ and
$q_\st = q{-}x_1P_1{-}x_2P_2$.

In averaged and weighted cross sections we will encounter contractions of hard 
and soft pieces like
\begin{equation}\begin{split}\label{sigma}
\Sigma(x_1,x_2,z_1,z_2,y)
&=\int d^2 p_{1\st}\,d^2 p_{2\st}\,d^2 k_{1\st}\,d^2 k_{2\st}\ 
\tr\big\{\Phi(x_1, p_{1\st})\Phi(x_2, p_{2\st})
\Delta(z_1, k_{1\st})\Delta(z_2, k_{2\st})
\,H\,H^\ast\,\big\}\\ 
&=\tr\big\{\,\Phi(x_1)\Phi(x_2)\Delta(z_1)\Delta(z_2)\,H\,H^\ast\,\big\}\ ,
\end{split}
\end{equation}
and
\begin{align}\label{sigmatrans}
\Sigma_\partial^\alpha(x_1,x_2,z_1,z_2,y)
&=\int d^2 p_{1\st}\,d^2 p_{2\st}\,d^2 k_{1\st}\,d^2 k_{2\st}\ 
q_\st^\alpha\,\tr\big\{\Phi(x_1, p_{1\st})\Phi(x_2, p_{2\st})
\Delta(z_1, k_{1\st})\Delta(z_2, k_{2\st})
\,H\,H^\ast\,\big\}\nonumber\\ 
&=\tr\big\{\,\big[\,\Phi_\partial^\alpha(x_1)\Phi(x_2)\Delta(z_1)\Delta(z_2) 
+\Phi(x_1)\Phi_\partial^\alpha(x_2)\Delta(z_1)\Delta(z_2)\nonumber\\
&\mbox{}\hspace{2cm}
\mbox{}-\Phi(x_1)\Phi(x_2)\Delta_\partial^\alpha(z_1)\Delta(z_2) 
-\Phi(x_1)\Phi(x_2)\Delta(z_1)\Delta_\partial^\alpha(z_2)\,\big]\,
H\,H^\ast\,\big\}\ .
\end{align}
These expressions for hadron-hadron scattering (and similar ones for
hadron-jet and jet-jet scattering) are schematic in the sense that the tracing
depends on the particular term in the sum of squared amplitudes, 
including both direct and interference
diagrams when the hard amplitude contains more than one contribution. 

In the case of hadron-hadron or hadron-jet cross sections one finds
averaged cross sections like
\begin{subequations}\label{hh-WeightedCrossSec}
\begin{align}
\langle\,d\sigma[h_1h_2]\,\rangle 
&=\int d\phi_2\ \frac{d\sigma[h_1h_2]}{d\phi_2}\nonumber\\
&=\frac{dx_{1\perp}\,dx_{2\perp}\,d\eta_1\,d\eta_2}{32\,\pi\,s}\,
\frac{d\phi_1}{2\pi}
\int \frac{dx_\perp}{x_\perp}\ 
\Sigma(x_1,x_2,z_1,z_2,y)\ ,\label{hh-average}\\
%%%%
\langle\,\tfrac{1}{2}\sin(\delta\phi)\,d\sigma[h_1h_2]\,\rangle
&=\int d\phi_2\ \tfrac{1}{2}\sin(\delta\phi)
\,\frac{d\sigma[h_1h_2]}{d\phi_2}\nonumber\\
&=\frac{dx_{1\perp}\,dx_{2\perp}\,d\eta_1\,d\eta_2}{32\,\pi\,s^{3/2}}\,
\frac{d\phi_1}{2\pi} \int \frac{dx_\perp}{x_\perp^2}\ 
e_{1\sN}{\cdot}\Sigma_{\partial}(x_1,x_2,z_1,z_2,y)\ .\label{hh-weight}
\end{align}
\end{subequations}
We would like to note that in Eq.~\eqref{hh-weight}, one is weighing with a 
dimensionless quantity which leads to a suppression with $1/\sqrt s$.
For jet-jet cross sections one has, in principle, the possibility to
access both perpendicular directions of $\Sigma_\partial^\alpha$,
assuming that $q = k_1{+}k_2$ is known accurately.
One could, then, weigh with $q_\st^\alpha$, in analogy to the Drell-Yan 
process~\cite{Tangerman:1994eh}. In that case one weighs with 
dimensionful quantities,
even if these are small momenta, and one does not get additional suppression
involving the hard scale.

\begin{figure}
\centering
\begin{minipage}{2cm}
\centering
\includegraphics[width=2cm]{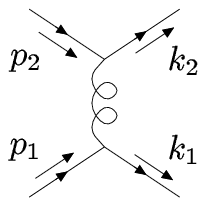}
(a)
\end{minipage}\hspace{0.5cm}
\begin{minipage}{2cm}
\centering
\includegraphics[width=2cm]{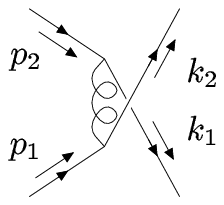}
(b)
\end{minipage}\hspace{0.5cm}
\begin{minipage}{2cm}
\centering
\includegraphics[width=2cm]{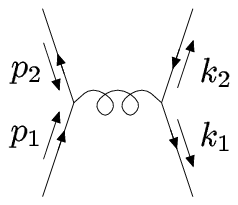}
(c)
\end{minipage}\hspace{0.5cm}
\begin{minipage}{2cm}
\centering
\includegraphics[width=2cm]{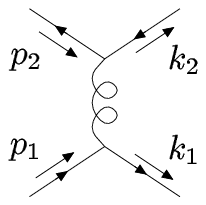}
(d)
\end{minipage}
\parbox{0.75\textwidth}{
\caption{\label{TUdiagrams}
Hard scattering amplitudes for quark-quark scattering:
(a) $t$-diagram, (b) $u$-diagram; 
quark-antiquark scattering:
(c) $s$-diagram, (d) $t$-diagram.}}
\end{figure}

These equations will be the starting point in the calculation of cross
sections. One needs to calculate the quantities in Eqs.~\eqref{sigma}
and~\eqref{sigmatrans}. These expressions involve hard scattering 
amplitudes and soft correlators $\Phi$ and 
$\Delta$, which are obtained as Fourier transforms of matrix elements 
of nonlocal combinations of quark and gluon fields. They are
parametrized in terms of distribution and fragmentation functions as
presented in appendices~\ref{CONSEQUENCES} and \ref{CONSEQUENCES2}.
In order to render the correlators color gauge invariant a gauge-link
connecting the fields is needed. In the diagrammatic calculation gauge-links
are explicitly found by taking into account, for each of the hadrons, 
the interactions of collinear gluons (polarizations along hadron momentum)
between the soft and hard parts. These give the well-known straight-line 
gauge-links along the lightcone for transverse momentum integrated 
correlators~\cite{Efremov:1978xm}, but they lead to nontrivial gauge-link 
paths for the TMD
correlators~\cite{Belitsky:2002sm,Boer:2003cm,Bomhof:2004aw}. 
The integration paths in the gauge-links $\mathcal U$ are process dependent, 
depending in particular on the hard partonic subprocesses.
We indicate this dependence by a superscript $\Phi^{[\mathcal U]}(x,p_\st)$.

The transverse momentum integrated correlator is a lightcone correlator with
a unique gauge-link, in which the path dependence disappears:
\begin{equation}
\Phi^{[\mathcal U]}(x)=\int d^2p_\st\ \Phi^{[\mathcal U]}(x,p_\st)=\Phi(x)\ .
\end{equation}
For the transverse moments of the correlators obtained after $p_\st$-weighting,
of which we only consider the simplest one, one finds two types of lightcone
correlators, a quark-quark matrix element $\Phi_\partial$
and a gluonic pole matrix element $\Phi_G$, 
where the latter is multiplied by a factor that depends on the gauge-link
\begin{equation}
\label{decomposition}
\Phi_\partial^{[\mathcal U]\alpha}(x) = 
\int d^2p_\st\ p_\st^\alpha\ \Phi^{[\mathcal U]}(x,p_\st) 
= \Phi_\partial^\alpha(x) + C_G^{[\mathcal U]}\,\pi\Phi_G^\alpha(x,x)\ .
\end{equation}
The gluonic pole matrix element, which contains the $T$-odd distribution
functions, was suggested in a slightly different
context by Qiu and Sterman~\cite{Qiu:1991wg,Qiu:1998ia} as the origin of 
single spin asymmetries. In processes like SIDIS with underlying hard process 
$\ell{+}q \rightarrow \ell{+}q$ and the DY process with underlying 
hard process $q{+}\bar q \rightarrow \ell{+}\bar\ell$, different 
gauge-link paths ${\mathcal U}^{[+]}$ and ${\mathcal U}^{[-]}$ appear.
In these processes the corresponding factors in Eq.~\eqref{decomposition} are 
simply $C_G^{[{\mathcal U}^{[\pm]}]} = \pm 1$. 

As was shown in Ref.~\cite{Bomhof:2004aw}, more complex integration paths enter in the gauge-links when other subprocesses are involved, 
such as the two-to-two (anti)quark subprocesses in this paper. 
Moreover, in general several diagrams enter
in the calculation. For instance, for quark-antiquark scattering both
$t$- and $s$-channel amplitudes (see Fig.~\ref{TUdiagrams}) can contribute,
$H = H_{q\bar q}^t{+}H_{q\bar q}^s$, whereas for quark-quark 
scattering we get $t$- and $u$-channel amplitudes, 
$H_{qq} = H_{qq}^t{+}H_{qq}^u$. 
In addition, one has to consider the diagrams
that produce the gauge-links needed to render the correlation functions
color gauge invariant. In these cases the gauge-links, in general, also differ
for the various terms 
appearing in the squared amplitude $HH^\dagger$ for a given partonic
subprocess.
For instance, in the scattering of two identical quarks 
the squared amplitude contains the terms 
$H_{qq}^tH_{qq}^{t\dagger}$, $H_{qq}^uH_{qq}^{u\dagger}$,
$H_{qq}^tH_{qq}^{u\dagger}$ and $H_{qq}^uH_{qq}^{t\dagger}$.
Details are explained in appendix~\ref{a}.

The results of the diagrammatic calculation for transverse momentum integrated
cross sections involving soft and hard parts can, in leading order, 
be recast in the form of a folding of the quark distribution and fragmentation 
functions appearing in the transverse momentum integrated correlators
$\Phi(x)$ and $\Delta(z)$ with hard partonic cross sections. 
For SIDIS one has a folding with the cross section for the hard process 
$\ell{+}q \rightarrow \ell{+}q$ and in the DY process a folding with the 
cross section for $q{+}\bar q \rightarrow \ell{+}\bar\ell$.
In hadron-hadron scattering one has several hard processes. An example is
$qq$ scattering with a cross section, in the case of identical 
quark flavors, of the form
\begin{equation}
\frac{d\hat\sigma_{qq\rightarrow qq}}{d\hat t}
= \sum_D 
\frac{d\hat\sigma_{qq\rightarrow qq}^{[D]}}{d\hat t}\ ,
\end{equation}
where the summation is over the different direct and interference 
contributions involving the $t$- and $u$-channel amplitudes.

A folding of distribution and fragmentation functions
is also possible for weighted cross sections. The cross sections involving 
the link-independent parts of the transverse moments
(i.e.\ $\Phi_\partial(x)$ and $\Delta_\partial(x)$) 
also lead to a folding with the normal partonic cross sections, 
just as for the integrated correlators $\Phi(x)$ and $\Delta(z)$.
However, for the contractions with the gluonic pole matrix elements
$\pi\Phi_G$ and $\pi\Delta_G$ the gauge-link dependence in the decomposition in Eq.~\eqref{decomposition} has important ramifications.
Expressing the asymmetries as a folding of universal,
one argument functions and a hard part requires a modification of the
hard partonic cross section by including the gauge-link dependent factors 
$C_G^{[{\mathcal U}]}$ in the various terms in this cross section. 
This is a convenient way of doing since the values of the factors $C_G$
depend on these terms.
For instance, in the example of unpolarized $qq$ scattering for 
identical flavors, the functions appearing in the parametrization of the gluonic pole matrix elements are
folded with the {\em gluonic pole cross section}
\begin{equation}
\frac{d\hat\sigma_{\widehat{gq}q\rightarrow qq}}{d\hat t}
= \sum_D C_G^{[\mathcal U(D)]}\,
\frac{d\hat\sigma_{qq\rightarrow qq}^{[D]}}{d\hat t}\ .
\end{equation}
The notation $\widehat{gq}$ emphasizes which quark field (in this case the
first one) is accompanied by a zero momentum gluon field in 
the correlator $\pi\Phi_G$. 
The parametrization of this correlator involves one-argument
distribution functions, which will appear folded with the gluonic pole
cross sections. 
At tree level often only one diagram contributes to the partonic cross section. 
In that case the gluonic pole cross section is simply proportional to the 
normal partonic cross section. 
For instance, the sign difference between
SIDIS and DY for the Sivers distribution function, a uniquely defined
function appearing in the parametrization of $\pi\Phi_G(x,x)$, comes from 
the factors $C_G^{[{\mathcal U}^{[\pm]}]} = \pm 1$ discussed above.
Instead of the folding with partonic cross sections, the Sivers function
is folded with the gluonic pole cross sections
\begin{subequations}
\begin{align}
\frac{d\hat\sigma_{\ell\widehat{gq}\rightarrow \ell q}}{d\hat t}
&= +\frac{d\hat\sigma_{\ell q\rightarrow \ell q}}{d\hat t}\ ,\\
\frac{d\hat\sigma_{\widehat{gq}\bar q\rightarrow \ell \bar \ell}}{d\hat t}
&= -\frac{d\hat\sigma_{q\bar q\rightarrow \ell \bar \ell}}{d\hat t}\ .
\end{align}
\end{subequations}
Although the gluonic pole cross sections should 
not be interpreted as true partonic cross sections, their
concept is convenient in order to get a simple folding expression involving the
one-argument functions appearing in the gluonic pole matrix elements
$\pi\Phi_G$ and $\pi\Delta_G$.
Moreover, they are easily obtained from the
terms in the hard partonic cross section without the inclusion of collinear gluon interactions between the hard and soft parts. 

In the following sections the formalism described above is applied to 
single spin asymmetries in inclusive two-hadron production, 
hadron-jet and jet-jet production in $p^\uparrow p$
scattering, for which the gluonic pole cross sections for
polarized (anti)quark scattering are also needed.

\section{
Single-spin asymmetries in inclusive hadron-hadron scattering\label{SSA}}
%$p{+}p\rightarrow\pi{+}\pi{+}X$}

As a reference we first consider the cross section for 2-particle inclusive 
hadron-hadron scattering. 
The explicit expression for the cross section in terms 
of the distribution and fragmentation functions can be obtained by inserting 
the parametrizations of the correlators, 
Eqs.~\eqref{QuarkCorr} and~\eqref{FragCorr}, 
into Eq.~\eqref{hh-average} and performing the required traces.
In this paper we restrict ourselves to the quark and antiquark scattering 
contributions.
Since the short-distance scattering subprocesses remain unobserved,
all partonic subprocesses that could contribute have to be taken into account.
This includes, for a realistic description, besides the (anti)quark 
contributions $qq{\rightarrow}qq$, $q\bar q{\rightarrow}q\bar q$ and 
$\bar q\bar q{\rightarrow}\bar q\bar q$, 
also contributions involving gluons, $qg{\rightarrow}qg$, 
$\bar q g{\rightarrow}\bar q g$, $gg{\rightarrow}gg$, 
$q\bar q{\rightarrow}gg$ and $gg{\rightarrow}q\bar q$ including their 
polarizations~\cite{Anselmino:1994tv,Anselmino:2002pd,Boer:2003tx,
D'Alesio:2004up,Anselmino:2004ky,Ma:2004tr,Schmidt:2005gv}.
However, the (anti)quark contributions suffice to illustrate how the 
inclusion of gauge-links leads to altered strengths of specific 
distribution or fragmentation functions. Contributions 
involving gluons can simply be added incoherently to the results 
presented here.
For the (anti)quark contributions to the averaged cross section one obtains
\begin{subequations}\label{Eq27}
\begin{align}
\langle\,d\sigma[h_1h_2]\,\rangle
&=dx_{1\perp}\,dx_{2\perp}\,d\eta_1\,d\eta_2\,\frac{d\phi_1}{2\pi}
\int\frac{dx_\perp}{x_\perp}
\nonumber\\
&\mspace{50mu}\times\,\frac{2\pi\,\alpha_\sS^2}{9\,\hat s}\,\bigg\{\ 
\big((1-y)^2+y^2\big)
\sum_{q,q'} f_1^q(x_1)\bar f_1^q(x_2)D_1^{q'}(z_1)\bar D_1^{q'}(z_2)\\
&\mspace{200mu}
+\frac{(1-y)^2+1}{y^2}
\sum_{q,q'} f_1^q(x_1)f_1^{q'}(x_2)D_1^q(z_1)D_1^{q'}(z_2)\displaybreak[0]\\
&\mspace{200mu}
+\frac{(1-y)^2+1}{y^2}
\sum_{q,q'}f_1^q(x_1)\bar f_1^{q'}(x_2)
D_1^q(z_1)\bar D_1^{q'}(z_2)\displaybreak[0]\\
&\mspace{200mu}
+\frac{2}{3}\frac{(1-y)^2}{y}
\sum_q f_1^q(x_1)\bar f_1^q(x_2)D_1^q(z_1)\bar D_1^q(z_2)\\
&\mspace{200mu}
-\frac{1}{3}\frac{1}{y(1-y)}
\sum_q f_1^q(x_1)f_1^q(x_2)D_1^q(z_1)D_1^q(z_2)\\
&\mspace{200mu}
+\big(\,\text{quark PDFs/FFs}\leftrightarrow\text{antiquark PDFs/FFs}\,\big)
\ \ \bigg\}%\nonumber\\
%&\mspace{300mu}
+\big(\,K_1\leftrightarrow K_2\,\big)\nonumber%\\
\end{align}
\end{subequations}
where the summation is over all quark flavors, including the case that $q=q'$.
In this expression $y$ is given by Eq.~\eqref{Y} and $\hat s$ is
$\hat s=x_\perp^2\,s\,\cosh^2[\frac{1}{2}(\eta_1{-}\eta_2)]$
= $x_\perp^2\,s/4y(1{-}y)$.
This result can be recast into a folding of the distribution and
fragmentation functions appearing in $\Phi(x)$ and $\Delta(z)$ and 
the elementary (anti)quark cross sections given in 
appendix~\ref{PARTONX}.
That is, expression~\eqref{Eq27} can be rewritten to
\begin{equation}
\langle\,d\sigma[h_1h_2]\,\rangle
= dx_{1\perp}\,dx_{2\perp}\,d\eta_1\,d\eta_2\,\frac{d\phi_1}{2\pi}
\int \frac{dx_\perp}{x_\perp}\ 
\sum_{q_1q_2q_3q_4}
f_1^{q_1}(x_1)f_1^{q_2}(x_2)\,\frac{\hat s}{2}
\,\frac{d\hat \sigma_{q_1q_2\rightarrow q_3q_4}}{d\hat t}\,
D_1^{q_3}(z_1)D_1^{q_4}(z_2)\ ,\label{UNPOLARIZED}
\end{equation}
where the summation is over all quark and antiquark flavors.
In the expressions above the momentum fractions are fixed by
$x_{1/2}=\tfrac{1}{2}x_\perp\big(\,e^{\pm\eta_1}{+}e^{\pm\eta_2}\,\big)$
and $z_i=x_{i\perp}/x_\perp$.

\subsection{
Hadron-hadron production in $p^\uparrow p$ scattering:
$p^\uparrow{+}p\rightarrow\pi{+}\pi{+}X$}

With only one of the hadrons polarized, any nonzero spin asymmetry
must involve at least one $T$-odd function. Restricting ourselves to
hadrons with spin $0$ and $\frac{1}{2}$, such functions do not show up in the 
transverse momentum integrated correlators $\Phi(x)$ and $\Delta(z)$. 
They do appear in the
parametrization of the matrix elements 
involved in the decomposition of the transverse moments of the correlators.
$T$-odd distribution functions only appear in the gluonic pole matrix element
$\pi\Phi_G$, while $T$-odd fragmentation functions can appear in both the
matrix elements $\Delta_\partial$ and
$\pi\Delta_G$~\cite{Boer:2003cm}.
Using the parametrizations for these functions, one can
calculate $e_{1\sN}{\cdot}\Sigma_\partial(x_1,x_2,z_1,z_2,y)$ and find
the expression for the weighted cross section using Eq.~\eqref{hh-weight}.
Considering only the (anti)quark contributions in 
$p^\uparrow{+}p\rightarrow\pi{+}\pi{+}X$ the resulting cross
section is explicitly given in 
appendix~\ref{diagrammaticresults}, including in each term explicitly
the factor $C_G^{[\mathcal U]}$ between braces $\{\,\cdot\,\}$.

The results can most conveniently be expressed as a folding of distribution
and fragmentation functions, now including one $T$-odd 
function and a {\em gluonic pole cross section}
\begin{subequations}\label{EXPR1}
\begin{align}
\langle\,\tfrac{1}{2}\sin&(\delta\phi)\,d\sigma[h_1h_2]\,\rangle\nonumber\\
&= dx_{1\perp}\,dx_{2\perp}\,d\eta_1\,d\eta_2\,
\frac{d\phi_1}{2\pi}\ \cos(\phi_1^S)
\int \frac{dx_\perp}{x_\perp}\ 
\nonumber\\
&\mspace{30mu}\times\bigg\{\ 
\frac{M_1}{x_\perp\,\sqrt s}
\sum_{q_1q_2q_3q_4}{f^{q_1}}_{1T}^{\perp(1)}(x_1)f_1^{q_2}(x_2)\,
\frac{\hat s}{2}
\,\frac{d\hat\sigma_{\widehat{gq}{}_1q_2\rightarrow q_3q_4}}{d\hat t}\,
D_1^{q_3}(z_1)D_1^{q_4}(z_2)\label{EXPR1a}\\
&\mspace{80mu}
+\frac{M_2}{x_\perp\,\sqrt s}
\sum_{q_1q_2q_3q_4}\,
h_1^{q_1}(x_1){h^{q_2}}_1^{\perp(1)}(x_2)\,
\frac{\hat s}{2}\,
\frac{d\Delta\hat\sigma_{q_1^\uparrow \widehat{gq}{}_2^\uparrow
\rightarrow q_3q_4}}{d\hat t}\,
D_1^{q_3}(z_1)D_1^{q_4}(z_2)\label{EXPR1b}\displaybreak[0]\\
&\mspace{80mu}
-\frac{M_{h_1}}{x_\perp\,\sqrt s}
\sum_{q_1q_2q_3q_4}
h_1^{q_1}(x_1)f_1^{q_2}(x_2)\,
\frac{\hat s}{2}
\,\frac{d\Delta\hat\sigma_{q_1^\uparrow q_2^{\phantom{\uparrow}}\rightarrow 
q_3^\uparrow q_4^{\phantom{\uparrow}}}}
{d\hat t}\,
H^{q_3}{}_1^{\perp(1)}(z_1)D_1^{q_4}(z_2)\ 
+\big(K_1{\leftrightarrow}K_2\big)\label{EXPR1c}\\
&\mspace{80mu}
-\frac{M_{h_1}}{x_\perp\,\sqrt s}
\sum_{q_1q_2q_3q_4}
h_1^{q_1}(x_1)f_1^{q_2}(x_2)\,
\frac{\hat s}{2}\,
\frac{d\Delta\hat\sigma_{q_1^\uparrow q_2^{\phantom{\uparrow}}
\rightarrow\widehat{gq}{}_3^\uparrow q_4^{\phantom{\uparrow}}}}{d\hat t}\,
\widetilde H^{q_3}{}_1^{\perp(1)}(z_1)D_1^{q_4}(z_2)\ 
+\big(K_1{\leftrightarrow}K_2\big)\ \bigg\}\label{EXPR1d}
\end{align}
\end{subequations}
where the summations run over all quark and antiquark flavors
and the angle $\phi_1^S$ is defined by $\phi_1^S = \phi_1{-}\phi_\sS$.
All non-vanishing partonic scattering cross sections and 
gluonic pole cross sections
are functions of $y$ or, equivalently, $\eta_1{-}\eta_2$ and those that
contribute to hadronic pion production are listed in appendix~\ref{PARTONX}.

We note the occurrence of {\em one} $T$-odd function in each of the terms 
in Eq.~\eqref{EXPR1}, the functions $f_{1T}^{\perp(1)}(x)$ and
$h_1^{\perp(1)}(x)$ coming from the gluonic pole matrix element
$\pi\Phi_G$, the function $H_1^{\perp(1)}(z)$ coming from the
link-independent correlator $\Delta_\partial$ and the function 
$\widetilde H_1^{\perp(1)}(z)$ coming from the gluonic pole matrix element
$\pi\Delta_G$. 
We would, once more, like to emphasize that for fragmentation both
$\Delta_\partial$ and $\pi\Delta_G$ contain $T$-odd functions
that could contribute to the Collins effect.
In Refs~\cite{Metz:2002iz,Collins:2004nx} it is argued that 
the Collins effect is universal.
This situation would occur if gluonic pole matrix elements 
in the case of fragmentation vanish,
in which case the function $\widetilde H_1^{\perp(1)}$ vanishes and all
$T$-odd effects come from the `universal' function $H_1^{\perp(1)}$.
The latter function appears folded with ordinary partonic cross sections.
However, in this paper we will allow for the gluonic pole matrix 
element for fragmentation and a nonvanishing function 
$\widetilde H_1^{\perp(1)}$, one reason being that there is no full 
agreement between the different model calculations concerning the 
universality of the Collins effect~\cite{Gamberg:2003eg,Amrath:2005gv}.

\subsection{
Hadron-jet production in $p^\uparrow p$ scattering:
$p^\uparrow{+}p\rightarrow\pi{+}\text{Jet}{+}X$}

We only take into account (anti)quark scattering processes
in the weighted scattering cross section for 
$p^\uparrow{+}p\rightarrow\pi{+}\text{Jet}{+}X$
with the pion and the jet approximately back-to-back in the perpendicular 
plane. This cross section can be obtained from the more involved 
two-particle inclusive scattering cross section~\eqref{EXPR} by taking 
$D_1^q(z_2)=\delta(z_2{-}1)\delta^{j_2q}
=x_\perp\,\delta(x_{2\perp}{-}x_\perp)\delta^{j_2q}$ and by 
letting all other fragmentation functions vanish.  
Here $\delta^{j_2q}$ is a delta function in flavor space,
indicating that the jet $j_2$ is produced by quark $q$.
The explicit expression using the diagrammatic approach is given in 
appendix~\ref{diagrammaticresults}. This can be recast into a form involving distribution and fragmentation functions folded with gluonic pole cross sections
\begin{subequations}
\begin{align}
\langle\,\tfrac{1}{2}\sin&(\delta\phi)\,d\sigma[h_1j_2]\,\rangle\nonumber\\
&=dx_{1\perp}\,dx_{2\perp}\,d\eta_1\,d\eta_2\,
\frac{d\phi_1}{2\pi}\ \cos(\phi_1^S)\nonumber\\
&\mspace{30mu}\times\bigg\{\ 
\frac{M_1}{x_{2\perp}\,\sqrt s}
\sum_{q_1q_2q_3q_4}{f^{q_1}}_{1T}^{\perp(1)}(x_1)f_1^{q_2}(x_2)\,
\frac{\hat s}{2}
\,\frac{d\hat\sigma_{\widehat{gq}{}_1q_2\rightarrow q_3q_4}}{d\hat t}\,
D_1^{q_3}(z_1)\\
&\mspace{80mu}
+\frac{M_2}{x_{2\perp}\,\sqrt s}
\sum_{q_1q_2q_3q_4}\,
h_1^{q_1}(x_1){h^{q_2}}_1^{\perp(1)}(x_2)\,
\frac{\hat s}{2}\,
\frac{d\Delta\hat\sigma_{q_1^\uparrow \widehat{gq}{}_2^\uparrow
\rightarrow q_3q_4}}{d\hat t}\,
D_1^{q_3}(z_1)\displaybreak[0]\\
&\mspace{80mu}
-\frac{M_{h_1}}{x_{2\perp}\,\sqrt s}
\sum_{q_1q_2q_3q_4}
h_1^{q_1}(x_1)f_1^{q_2}(x_2)\,
\frac{\hat s}{2}
\,\frac{d\Delta\hat\sigma_{q_1^\uparrow q_2^{\phantom{\uparrow}}\rightarrow 
q_3^\uparrow q_4^{\phantom{\uparrow}}}}
{d\hat t}\,
H^{q_3}{}_1^{\perp(1)}(z_1)\\
&\mspace{80mu}
-\frac{M_{h_1}}{x_{2\perp}\,\sqrt s}
\sum_{q_1q_2q_3q_4}
h_1^{q_1}(x_1)f_1^{q_2}(x_2)\,
\frac{\hat s}{2}\,
\frac{d\Delta\hat\sigma_{q_1^\uparrow q_2^{\phantom{\uparrow}}
\rightarrow \widehat{gq}{}_3^\uparrow q_4^{\phantom{\uparrow}}}}{d\hat t}\,
\widetilde H^{q_3}{}_1^{\perp(1)}(z_1)\ \bigg\}
\end{align}
\end{subequations}

\subsection{
Jet-jet production in $p^\uparrow p$ scattering:
$p^\uparrow{+}p\rightarrow\text{Jet}{+}\text{Jet}{+}X$}

We only take into account (anti)quark scattering processes 
in the weighted scattering cross section for 
$p^\uparrow{+}p\rightarrow\text{Jet}{+}\text{Jet}{+}X$ 
with approximately back-to-back jets in the perpendicular plane.
As argued, in principle one can construct azimuthal asymmetries
that give access to $\Sigma_\partial^\alpha(x_1,x_2,y)$,
by weighting with the small momentum $q_\st^\alpha$. However, 
this requires accurate determination of the jet momenta $k_1$ and $k_2$. 
Here we only present the cross section obtained by weighting with
$\sin(\delta\phi)$, which can be obtained from the more involved two-particle
inclusive process~\eqref{EXPR} by taking 
$D_1^q(z_i)=\delta(z_i{-}1)\delta^{j_iq}
=x_\perp\,\delta(x_{i\perp}{-}x_\perp)\delta^{j_iq}$ 
and by letting all other fragmentation functions vanish.  
Casting the result from the diagrammatic approach (given explicitly
in appendix~\ref{diagrammaticresults}) in the form of a folding, 
one obtains
\begin{subequations}
\begin{align}
\langle\,\tfrac{1}{2}\sin&(\delta\phi)\,d\sigma[j_1j_2]\,\rangle\nonumber\\
&=dx_{1\perp}\,dx_{2\perp}\,d\eta_1\,d\eta_2\,
\frac{d\phi_1}{2\pi}\ \cos(\phi_1^S)\ \delta(x_{1\perp}{-}x_{2\perp})
\nonumber\\
&\mspace{30mu}\times\bigg\{\ 
\frac{M_1}{\sqrt s}
\sum_{q_1q_2q_3q_4}{f^{q_1}}_{1T}^{\perp(1)}(x_1)f_1^{q_2}(x_2)\,
\frac{\hat s}{2}
\,\frac{d\hat\sigma_{\widehat{gq}{}_1q_2\rightarrow q_3q_4}}{d\hat t}\\
&\mspace{80mu}
+\frac{M_2}{\sqrt s}
\sum_{q_1q_2q_3q_4}\,
h_1^{q_1}(x_1){h^{q_2}}_1^{\perp(1)}(x_2)\,
\frac{\hat s}{2}\,
\frac{d\Delta\hat\sigma_{q_1^\uparrow \widehat{gq}{}_2^\uparrow
\rightarrow q_3q_4}}{d\hat t}\ \bigg\}
\end{align}
\end{subequations}

\section{Summary and conclusions} 

In this paper we have used the diagrammatic approach at tree-level to 
derive expressions for single transverse-spin asymmetries in 2-particle
inclusive hadron-hadron collisions. The final states considered are
hadron-hadron, hadron-jet and jet-jet, which are approximately back-to-back
in the plane perpendicular to the incoming hadrons. The single spin
asymmetries require the inclusion of transverse momentum dependence
for the partons. We have assumed factorization to hold in this
treatment of TMD effects although it is, 
at present, certainly not clear whether such a factorization holds for 
hadron-hadron scattering processes with explicitly TMD correlators. We have
limited ourselves to the first transverse moments obtained by weighting
linearly with the transverse momentum. These transverse moments show
up in azimuthal asymmetries.

While single-spin asymmetries generated by
fragmentation processes, in which one can have $T$-odd fragmentation functions,
are well-known, the single-spin asymmetries connected with
initial state hadrons are more subtle. Within the diagrammatic
approach $T$-odd effects for transverse momentum dependent distribution 
functions are attributed to the structure of the integration path in the gauge-link.
This path depends on the specific hard process in which the correlator
is used, explaining, for instance, the appearance of the Sivers function
$f_{1T}^{\perp(1)}$ with opposite signs in SIDIS and 
DY~\cite{Brodsky:2002cx,Collins:2002kn,Brodsky:2002rv}. 
In the transverse moments of quark and antiquark correlators the effect
of the gauge-link appears via the gluonic pole matrix element, which 
in the case of distributions is a $T$-odd matrix element giving 
rise to single spin asymmetries~\cite{Qiu:1991wg,Qiu:1998ia}.
In this paper we show how the effects of the gauge-link appear as 
factors $C_G^{[\mathcal U]}$, which determine the strengths with which 
the gluonic pole matrix elements occur. This is a generalization
of the factors $\pm 1$ appearing in SIDIS and DY.
The fact that these strengths are determined by the hard parts makes it 
convenient to absorb them in so-called gluonic pole cross sections. 
Just as the transverse momentum averaged cross sections can, in leading order,
be written as a folding of universal distribution and fragmentation functions
and a hard partonic cross section,
the single spin asymmetries can be written as a folding of universal
distribution and fragmentation functions (involving one $T$-odd function) and a gluonic pole cross section. 

In our approach we allow for two possible mechanisms to produce
single spin asymmetries in the case of fragmentation. This implies 
that in the two matrix elements in which the transverse moments can be 
decomposed, i.e.\ the link-independent part
$\Delta_\partial$ and the gluonic pole matrix element $\pi\,\Delta_G$,
one has both $T$-even and $T$-odd effects.
For the Collins effect in fragmentation, it
leads to two independent functions $H_1^{\perp(1)}$ and 
$\widetilde{H}_1^{\perp(1)}$, the latter appearing in the parametrization of the gluonic pole matrix element. Having different linear combinations of these 
functions in SIDIS and electron-positron annihilation spoils the
comparison of the Collins effect in these processes. 
In hadron-hadron collisions we find
other linear combinations of the two functions. 
If fragmentation functions are universal, as is argued in
Refs.~\cite{Metz:2002iz,Collins:2004nx}, the tilde function 
$\widetilde{H}_1^{\perp(1)}$ (and the gluonic pole matrix element
for fragmentation) vanishes. In that case only the contribution
from $H_1^{\perp(1)}$ remains.

Our results, including the strengths of the gluonic pole
matrix elements differ from those of earlier calculations in
which the effects of the gauge-links have been omitted. 
However, these effects can easily be incorporated by using the 
gluonic pole cross sections instead of the normal hard
partonic cross sections. 
%As an example, we have compared
%$d\hat\sigma_{\widehat{gq}q\rightarrow qq}$ with 
%$d\hat\sigma_{qq\rightarrow qq}$ in Fig.~(figure).

We have restricted ourselves to a particular single spin asymmetry in hadron-hadron scattering where the asymmetry arises from the deviation from
the back-to-back appearance of the produced hadrons/jets in the perpendicular plane. 
This situation was discussed in Ref.~\cite{Boer:2003tx}
without inclusion of the effects of gauge-links. 
Although experimentally more challenging,
the 2-particle inclusive case is easier to analyze than the
1-particle inclusive case, where large single
spin asymmetries are observed, but where
subleading transverse momentum averaged $T$-odd fragmentation 
functions~\cite{Jaffe:1993xb} will also contribute.
In principle the diagrammatic approach allows for 
inclusion of these contributions. Furthermore, the methods
used in this paper to include the $T$-odd, transverse momentum
dependent effects in (anti)quark contributions, which are crucial to treat
single spin asymmetries in hadron-hadron scattering, can be extended 
to include the gluonic contributions as well as to treat various 
$T$-even double spin asymmetries.

\appendix

\section{quark correlators and gauge-links\label{a}}

The starting point for the structure of the hadron$\rightarrow$quark 
transition is the quark correlator
$\Phi(p)\equiv\Phi(p;P,S)$~\cite{Soper:1976jc,Collins:1981uw}
\begin{equation}
\Phi_{ij}(p;P,S)=\int\frac{d^4\xi}{(2\pi)^4}\ e^{ip\cdot\xi}
\langle P,S|\,\overline\psi_j(0)\psi_i(\xi)\,|P,S\rangle\ .
\end{equation}
Similarly, one has for the quark$\rightarrow$hadron transition 
the fragmentation correlator $\Delta(k)\equiv\Delta(k;K,S)$,
\begin{equation}
\Delta_{ij}(k;K,S_h)=\int\frac{d^4\xi}{(2\pi)^4}\ e^{ik\cdot\xi}
\langle0|\,\psi_i(\xi)\,a_h^\dagger a_h^{\phantom{\dagger}}\,
\overline\psi_j(0)\,|0\rangle\ ,
\end{equation}
with
\begin{equation}
a_h^\dagger a_h^{\phantom{\dagger}}
=\ \int\mspace{-26mu}\sum\nolimits_X\frac{d^3P_X}{(2\pi)^32E_X}\ 
|P_X;K,S_h\rangle\langle P_X;K,S_h|\ .
\end{equation}
In the description of hard scattering processes we need
the quark correlator and the fragmentation correlator 
integrated over, at least, the partonic momentum component
$p{\cdot}P$. This leaves the TMD correlator
\begin{equation}
\Phi(x,p_\st) =\int d(p{\cdot}P)\ \Phi(p)\ .
\end{equation}
Integrating the TMD correlator over or weighing it with the transverse momentum $p_\st$, we
obtain
\begin{subequations}
\begin{align}
\Phi(x)
&=\int d^2p_\st\ \Phi(x,p_\sT)\ ,\\
\Phi_\partial^\alpha(x)
&=\int d^2p_\st\ p_\st^\alpha\ \Phi(x,p_\sT)\ .
\end{align}
\end{subequations}
One finds similar expressions for the fragmentation correlator.
Analogous to the above one can write down the antiquark correlator 
$\overline\Phi$ 
describing the hadron$\rightarrow$antiquark transition and the antiquark 
fragmentation correlator $\overline\Delta$ describing the 
antiquark$\rightarrow$hadron transition.

\begin{figure}
\centering
\begin{minipage}[t]{2.2cm}
\centering
\includegraphics[width=2.2cm]{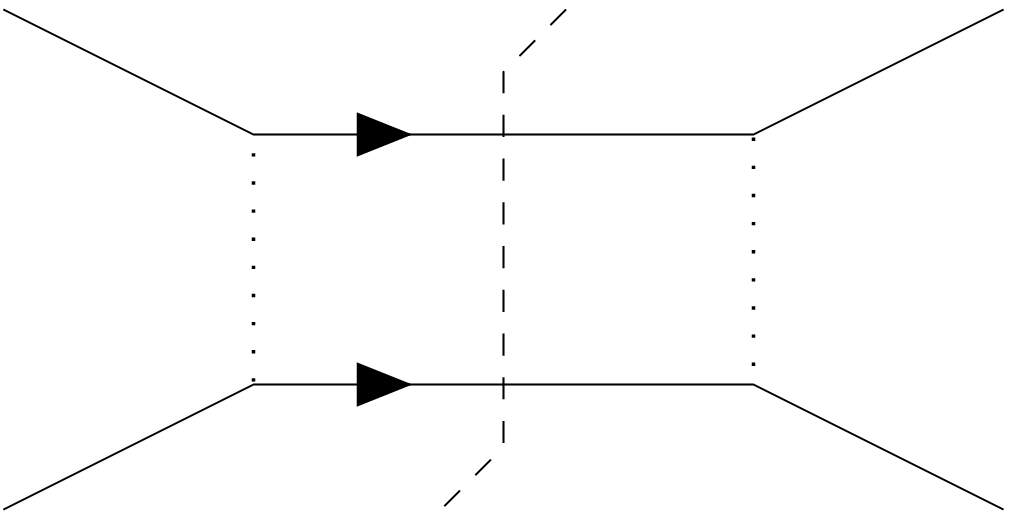}
\begin{equation*}
\frac{\tr(\mathcal U^{[\Box]})}{N_c}\mathcal U^{[+]}\ \,
\end{equation*}
\end{minipage}\hspace{3mm}
\begin{minipage}[t]{2.2cm}
\centering
\includegraphics[width=2.2cm]{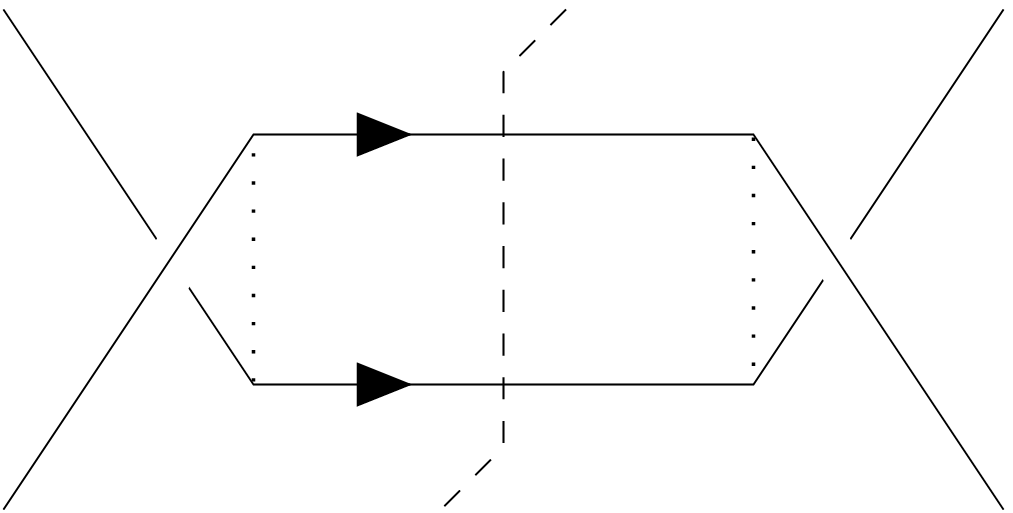}
\begin{equation*}
\frac{\tr(\mathcal U^{[\Box]})}{N_c}\mathcal U^{[+]}\ \,
\end{equation*}
\end{minipage}\hspace{3mm}
\begin{minipage}[t]{2.2cm}
\centering
\includegraphics[width=2.2cm]{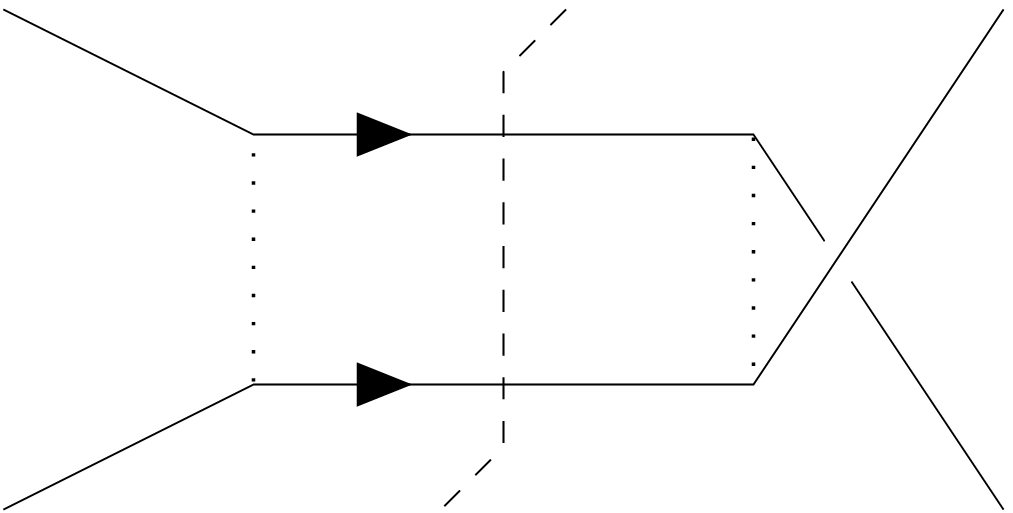}\\[-6pt]
\begin{equation*}
\mathcal U^{[\Box]}\mathcal U^{[+]}\quad\ 
\end{equation*}
\end{minipage}\hspace{3mm}
\begin{minipage}[t]{2.2cm}
\centering
\includegraphics[width=2.2cm]{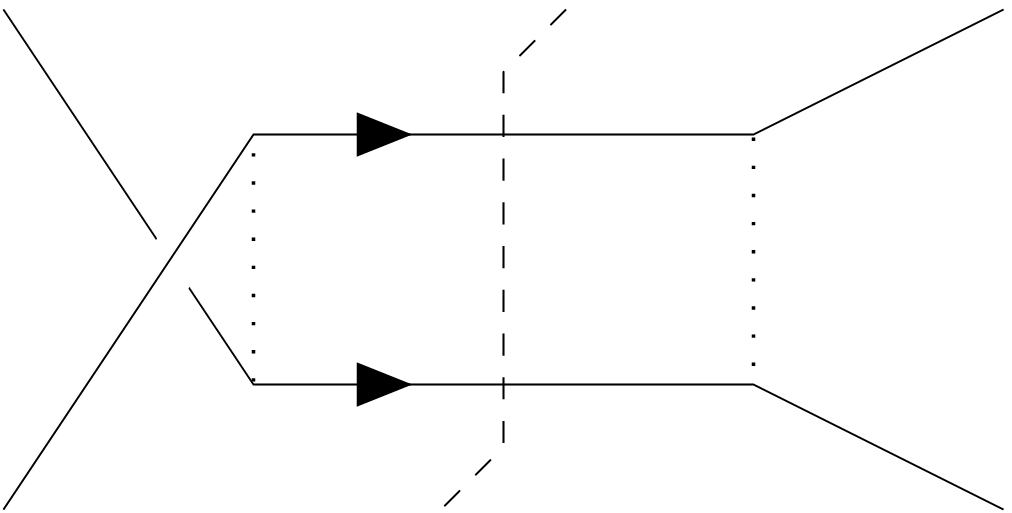}\\[-6pt]
\begin{equation*}
\mathcal U^{[\Box]}\mathcal U^{[+]}\quad\ 
\end{equation*}
\end{minipage}\\[4mm]
%%%%
\begin{minipage}[t]{2.2cm}
\centering
\includegraphics[width=2.2cm]{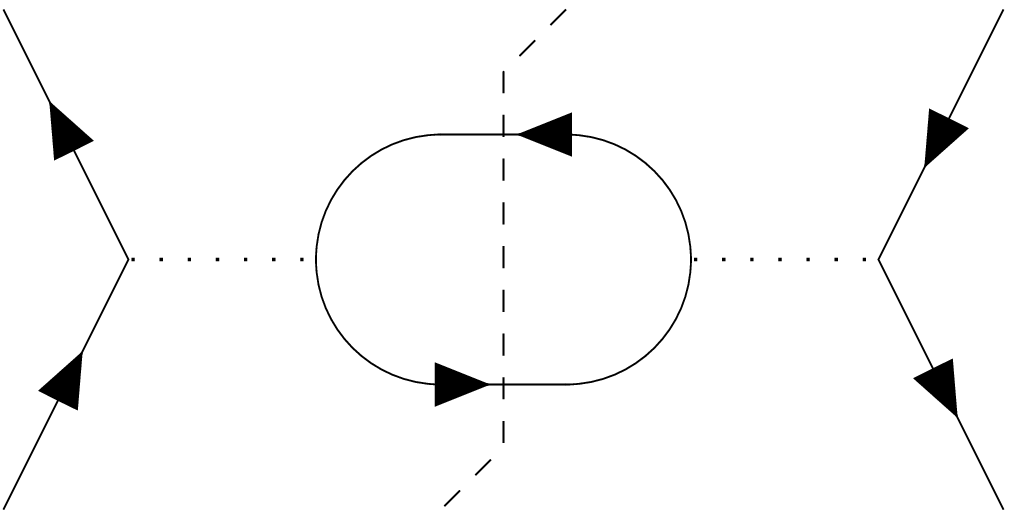}\\[-6pt]
\begin{equation*}
\mathcal U^{[-]}
\end{equation*}
\end{minipage}\hspace{3mm}
\begin{minipage}[t]{2.2cm}
\centering
\includegraphics[width=2.2cm]{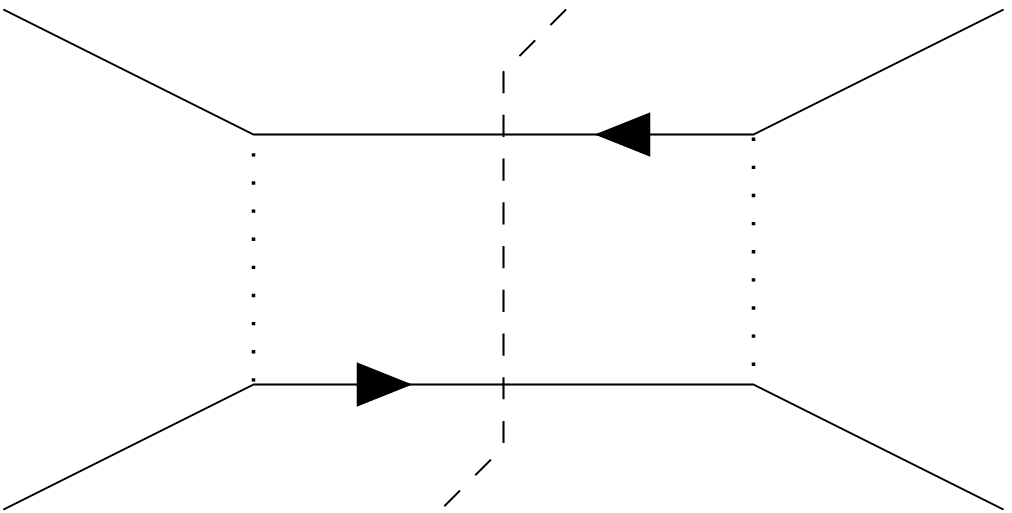}
\begin{equation*}
\frac{\tr(\mathcal U^{[\Box]\dagger})}{N_c}\mathcal U^{[+]}\ 
\end{equation*}
\end{minipage}\hspace{3mm}
\begin{minipage}[t]{2.2cm}
\centering
\includegraphics[width=2.2cm]{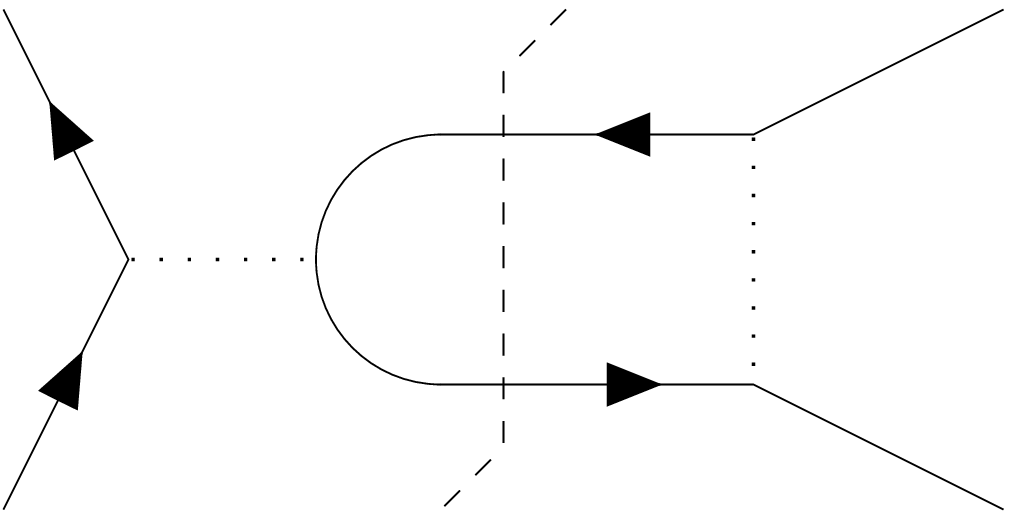}\\[-6pt]
\begin{equation*}
\mathcal U^{[-]}
\end{equation*}
\end{minipage}\hspace{3mm}
\begin{minipage}[t]{2.2cm}
\centering
\includegraphics[width=2.2cm]{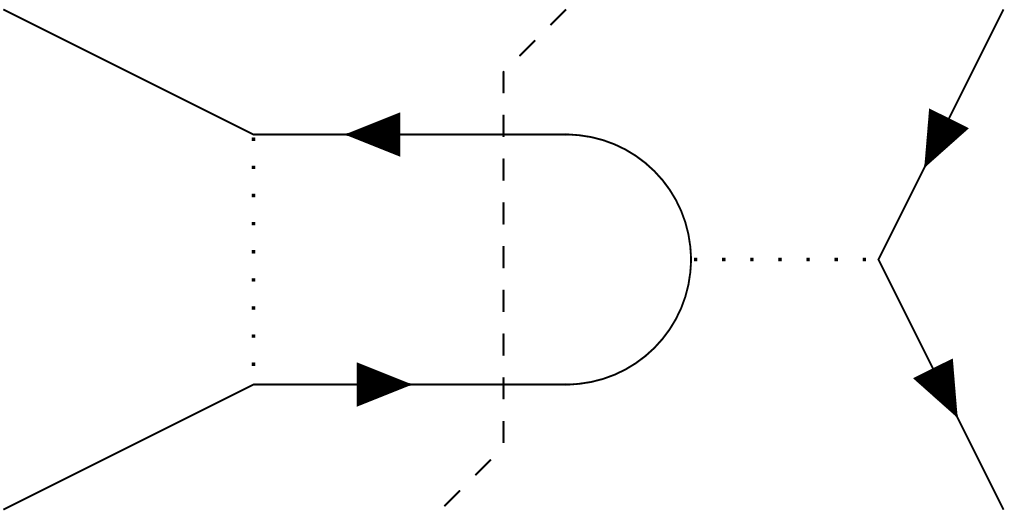}\\[-6pt]
\begin{equation*}
\mathcal U^{[-]}
\end{equation*}
\end{minipage}\hspace{3mm}
\parbox{0.7\textwidth}{
\caption{Gauge-links entering in the correlator for the lower-left incoming
quark for a hard two-fermion scattering process without exchange of charge.
Top: quark-quark scattering; 
bottom: quark-antiquark scattering.\label{QED}}}
\end{figure}

To obtain properly gauge invariant correlators,
gauge-links connecting the parton fields in the matrix elements are needed.
The general structure of the gauge-links is
$\mathcal U^{[C]}(0,\xi){=}\mathcal P\exp[\,-ig\int_CA(z){\cdot}dz\,]$,
where the integration path $C$ runs from 0 to $\xi$.
Here $A$ is the gauge field and $\mathcal P$ is the path-ordering operator.
The integration paths can be calculated by resumming all collinear gluon interactions between the soft and hard parts.
Consequently, for the TMD correlators they depend on the process in which they occur.
The gauge-links appearing in the quark correlator 
in a two-fermion hard scattering process with uncharged boson exchange,
such as in QED, are readily calculated by considering the flow of the fermion 
lines~\cite{Bomhof:2004aw}.
The results from that reference are given in Fig.~\ref{QED}.
Explicitly, we encounter the link structures 
\begin{align}
\mathcal U^{[\pm]}
&=U_{[(0^-,\boldsymbol0_T),(\pm\infty^-,\boldsymbol0_T)]}^-
U_{[(\pm\infty^-,\boldsymbol0_T),(\pm\infty^-,\boldsymbol\infty_T)]}^T
U_{[(\pm\infty^-,\boldsymbol\infty_T),(\pm\infty^-,\boldsymbol\xi_T)]}^T
U_{[(\pm\infty^-,\boldsymbol\xi_T),(\xi^-,\boldsymbol\xi_T)]}^-\ ,\label{GL}\\
\mathcal U^{[\Box]}&=\mathcal U^{[+]}\mathcal U^{[-]\dagger}\ ,
\end{align}
which are build up from the gauge-links along straight lines
\begin{align}
U_{[a,b]}^-
=\mathcal P\exp\Big[-ig\int_a^bdz\ n\cdot A(z)\,\Big]\ ,\qquad\text{and}\qquad
U_{[a,b]}^T
=\mathcal P\exp\Big[-ig\int_a^bdz_\st\cdot A_\st(z)\,\Big]\ .
\end{align}
The gauge-links in the scattering of two colored fermions in QCD can be 
obtained from those in Fig.~\ref{QED} by accounting for the flow of color 
charge, using well-known QCD rules for color flow such as
\begin{equation}
\parbox{1cm}{
\includegraphics[width=1cm]{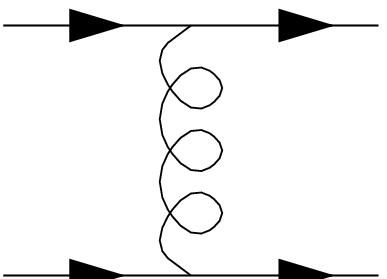}}
=T_F\left(\ \ 
\parbox{1cm}{
\includegraphics[width=1cm]{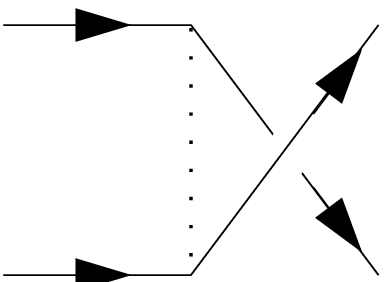}}
-\frac{1}{N_c}\ 
\parbox{1cm}{
\includegraphics[width=1cm]{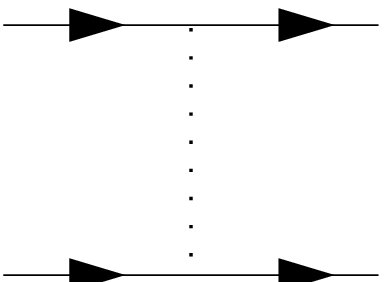}}
\ \ \right)\ .
\end{equation}
For example, the $tt^*$-channel of quark-quark scattering can be 
decomposed in this way giving
\begin{equation}
\parbox{2cm}{
\includegraphics[width=2cm]{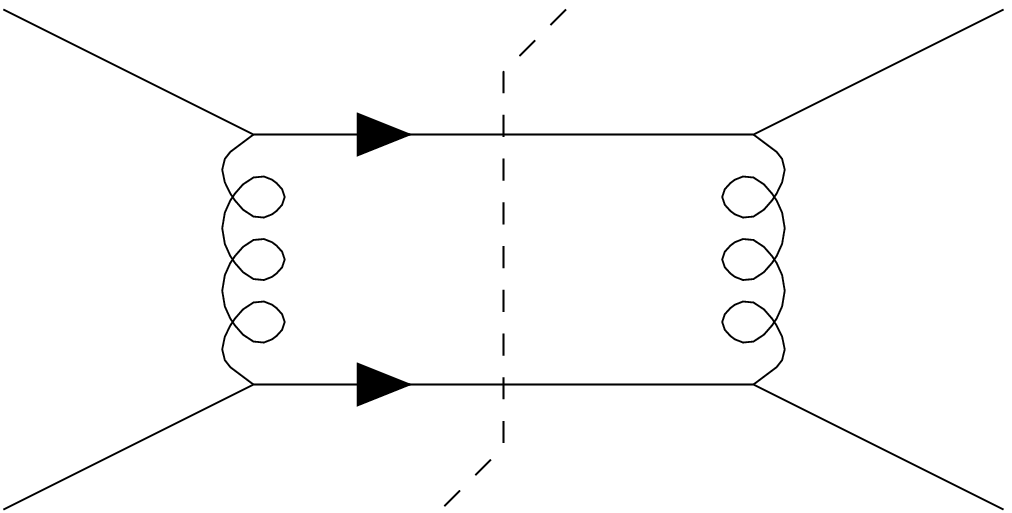}}
=T_F^2\left(\ 
\parbox{2cm}{
\includegraphics[width=2cm]{Figures/UUdiag}}
-\frac{1}{N_c}\ 
\parbox{2cm}{
\includegraphics[width=2cm]{Figures/UTdiag}}
-\frac{1}{N_c}\ 
\parbox{2cm}{
\includegraphics[width=2cm]{Figures/TUdiag}}
+\frac{1}{N_c^2}\ 
\parbox{2cm}{
\includegraphics[width=2cm]{Figures/TT2diag}}
\ \right)\ .
\end{equation}
The gauge-link of this diagram can be obtained by replacing each diagram 
on the r.h.s.\ with the corresponding QED gauge-link as given by 
Fig.~\ref{QED} and factoring out the overall color factor of the QCD diagram.
The overall color factor of the gluon exchange diagram on the l.h.s.\ is 
$(\tr[t^at^b])^2{=}T_F^2(N_c^2{-}1)$, which can also be obtained by
tracing the color flow in all diagrams on the r.h.s.
This color factor does not enter in the gauge-link,
but in the evaluation of the diagram itself and is included in the 
hard amplitudes that will be used in the calculations in
appendix~\ref{diagrammaticresults}.
Accounting for the additional factors $\tr(\openone)=N_c$ that are 
obtained for each color loop, one obtains the gauge-link
\begin{align*}
T_F^2(N_c^2-1)\times\mathcal U_{qq}^{[tt^*]}
&=T_F^2\Big\{\,
N_c^2{\times}\frac{\tr(\mathcal U^{[\Box]})}{N_c}\mathcal U^{[+]}
-\mathcal U^{[\Box]}\mathcal U^{[+]}
-\mathcal U^{[\Box]}\mathcal U^{[+]}
+\frac{\tr(\mathcal U^{[\Box]})}{N_c}\mathcal U^{[+]}\,\Big\}\ .
\end{align*}
The other diagrams can be calculated analogously.
For quark-quark scattering we obtain:
\begin{subequations}\label{QuarkQuark}
\begin{align}
\mathcal U_{qq}^{[tt^*]}
&=\mathcal U_{qq}^{[uu^*]}
=\frac{1}{N_c^2-1}\Big\{\,
(N_c^2+1)\,\frac{\tr(\mathcal U^{[\Box]})}{N_c}\mathcal U^{[+]}
-2\,\mathcal U^{[\Box]}\mathcal U^{[+]}\,\Big\}\ ,\label{QuarkQuarkA}\\
\mathcal U_{qq}^{[tu^*]}
&=\mathcal U_{qq}^{[ut^*]}
=\frac{N_c}{N_c^2-1}\Big\{\,
2N_c\,\frac{\tr(\mathcal U^{[\Box]})}{N_c}\mathcal U^{[+]}
-\frac{N_c^2+1}{N_c}\,\mathcal U^{[\Box]}\mathcal U^{[+]}\,\Big\}\ ,
\end{align}
\end{subequations}
and for quark-antiquark scattering:
\begin{subequations}\label{QuarkAntiquark}
\begin{align}
\mathcal U_{q\overline q}^{[ss^*]}
&=\frac{1}{N_c^2-1}\Big\{\,
N_c^2\,\frac{\tr(\mathcal U^{[\Box]\dagger})}{N_c}\mathcal U^{[+]}
-\mathcal U^{[-]}\,\Big\}\ ,\\
%%%%
\mathcal U_{q\overline q}^{[tt^*]}
&=\frac{1}{N_c^2-1}\Big\{\,
\frac{\tr(\mathcal U^{[\Box]\dagger})}{N_c}\mathcal U^{[+]}
+(N_c^2-2)\,\mathcal U^{[-]}\,\Big\}\ ,\\
%%%%
\mathcal U_{q\overline q}^{[st^*]}
&=\mathcal U_{q\overline q}^{[ts^*]}
=\frac{N_c}{N_c^2-1}\Big\{\,
N_c\,\frac{\tr(\mathcal U^{[\Box]\dagger})}{N_c}\mathcal U^{[+]}
-\frac{1}{N_c}\,\mathcal U^{[-]}\,\Big\}\ .
\end{align}
\end{subequations}
These are the gauge-link operators that enter between the quark fields 
in the correlator of the incoming quark:
\begin{equation}\label{LINKQuarkCorr}
\Phi^{[\mathcal U]}(x,p_\sT;P,S)
=\int\frac{d(\xi{\cdot}P)}{2\pi}\frac{d^2\xi_\sT}{(2\pi)^2}\ e^{ip\cdot\xi}
\langle P,S|\,\overline\psi(0)\,\mathcal U(0,\xi)\,\psi(\xi)\,|P,S\rangle\ .
\end{equation}
The gauge-links that enter in the quark-fragmentation correlators are
the time-reversed ones as compared to those in the quark-correlators. 
That is, a $\mathcal U^{[+]}$ in the quark-correlator corresponds to a 
$\mathcal U^{[-]}$ in the fragmentation correlator and a 
$\mathcal U^{[\Box]}$ to a $\mathcal U^{[\Box]\dagger}$, etc.
The gauge-links that enter in the antiquark-correlators $\overline\Phi$ 
and $\overline \Delta$ are the hermitian conjugates of the gauge-links 
in the quark-correlators $\Phi$ and $\Delta$ of the corresponding diagrams.

We note that for the fragmentation correlators the gauge-links 
are all split up in parts, parts connecting to the field at $\xi$ 
and others to the field at $0$.
Taking as an example the quark fragmentation correlators 
with the gauge-links $\mathcal U^{[\pm]}$, one has (compare with Eq.~\eqref{GL})
\begin{equation}\begin{split}\label{QuarkFrag}
\Delta^{[\pm]}(z,k_\sT;K,S)
=\int\frac{d(\xi{\cdot}K)}{2\pi}\frac{d^2\xi_\sT}{(2\pi)^2}\ e^{ik\cdot\xi}
\langle0|\,
&U_{[(\pm\infty^-,\boldsymbol\infty_T),(\pm\infty^-,\boldsymbol\xi_T)]}^T
U_{[(\pm\infty^-,\boldsymbol\xi_T),(\xi^-,\boldsymbol\xi_T)]}^-\,
\psi(\xi)\,\\
&\times 
a_h^\dagger a_h^{\phantom{\dagger}}\,\overline\psi(0)\,
U_{[(0^-,\boldsymbol0_T),(\pm\infty^-,\boldsymbol0_T)]}^-
U_{[(\pm\infty^-,\boldsymbol0_T),(\pm\infty^-,\boldsymbol\infty_T)]}^T\,
|0\rangle\ .
\end{split}\end{equation}

\section{
consequences of gauge-links for distribution functions\label{CONSEQUENCES}}

The gauge-link has important consequences for the parametrizations of the 
correlator in $T$-even and $T$-odd functions. 
We start with the link structures enumerated in Fig.~\ref{QED}.  
The TMD correlators are link dependent. We write
$\Phi^{[\pm]}$ for the correlators with gauge-links $\mathcal U^{[\pm]}$,
$\Phi^{[\Box+]}$ for $\mathcal U^{[\Box]}\mathcal U^{[+]}$,
$\Phi^{[(\Box)+]}$ for 
$\frac{1}{N_c}\tr(\mathcal U^{[\Box]})\mathcal U^{[+]}$ and
$\Phi^{[(\Box^\dagger)+]}$ for 
$\frac{1}{N_c}\tr(\mathcal U^{[\Box]\dagger})\mathcal U^{[+]}$.
We will also need the transverse momentum integrated
correlators $\Phi_D^\alpha(x)$ and $\Phi_G^\alpha(x,x{-}x')$
\begin{gather}
\Phi_D^\alpha(x)
=\int\frac{d(\xi{\cdot}P)}{2\pi}\ e^{ix(\xi\cdot P)}
\langle P,S|\,\overline\psi(0)\,U_{[0,\xi]}^{-}\,iD^\alpha(\xi)\,
\psi(\xi)\,|P,S\rangle\,\big\rfloor_{\text{LC}}\ ,\\
%%%%
\Phi_G^\alpha(x,x{-}x')
=\int\frac{d(\xi{\cdot}P)}{2\pi}\frac{d(\eta{\cdot}P)}{2\pi}\ 
e^{i(x-x')(\xi\cdot P)}e^{ix'(\eta\cdot P)}
\langle P,S|\,\overline\psi(0)\,U_{[0,\eta]}^{-}\,gG^{n\alpha}(\eta)\,
U_{[\eta,\xi]}^-\,\psi(\xi)\,|P,S\rangle\,\big\rfloor_{\text{LC}}\ ,
\end{gather}
which are set on the lightcone (LC) where $\xi\cdot n=\xi_\st=0$ and 
$\eta\cdot n=\eta_\st=0$.
We have also used the shorthand notation 
$G^{n\alpha}=g_{\mu\nu}G^{\mu\alpha}n^\nu$ for the field strength tensor.
In terms of these the weighted correlators can be written as
\begin{subequations}\label{ToddCors}
\begin{gather}
\Phi_\partial^{[\pm]\alpha}(x)
=\Phi_D^\alpha(x)
-\int dx'\ \frac{i}{x'\mp i\epsilon}\,\Phi_G^\alpha(x,x{-}x')
=\Phi_\partial^\alpha(x) \pm \pi\Phi_G^\alpha(x,x)\ ,\label{ToddCorsA}\\
%%%%
\Phi_\partial^{[\Box+]\alpha}(x)
=\Phi_D^\alpha(x)
-\int dx'\ 
\Big\{\,\frac{i}{x'-i\epsilon}-2\pi\delta(x')\,\Big\}\,
\Phi_G^\alpha(x,x{-}x')
=\Phi_\partial^\alpha(x)
+3\pi\Phi_G^\alpha(x,x)\ ,\label{ToddCorsB}\\
%%%%
\Phi_\partial^{[(\Box)+]\alpha}(x)
=\Phi_\partial^{[(\Box^\dagger)+]\alpha}(x)
=\Phi_D^\alpha(x)
-\int dx'\ \frac{i}{x'-i\epsilon}\,\Phi_G^\alpha(x,x{-}x')
=\Phi_\partial^\alpha(x)
+\pi\Phi_G^\alpha(x,x)\ ,\label{ToddCorsC}
\end{gather}
\end{subequations}
where $\Phi_\partial$ without link index refers to
\be
\Phi_\partial^\alpha(x)
=\Phi_D^\alpha(x)
-\int dx'\ P\frac{i}{x'}\ \Phi_G^\alpha(x,x{-}x')\ .\label{E4}\\
\ee
The decomposition in Eq.~\eqref{ToddCors} is useful because 
time reversal symmetry implies that the
correlator $\Phi_\partial$ only contains $T$-even functions, 
while $\Phi_G$ only contains $T$-odd functions.

The correlators encountered in $p^\uparrow p\rightarrow\pi\pi X$ are readily obtained from the results above and can also be decomposed in terms of
$\Phi_\partial$ and $\Phi_G$.
For instance, for the $tt^*$-channel in $qq$ scattering 
we get from Eq.~\eqref{QuarkQuarkA}
\begin{align*}
\Phi_\partial^{[tt^*]\alpha}(x)
&=\frac{1}{N_c^2-1}\Big\{\,(N_c^2+1)\Phi_\partial^{[(\Box)+]\alpha}(x)
-2\,\Phi_\partial^{[\Box+]\alpha}(x)\,\Big\}\\
&=\frac{1}{N_c^2-1}\big\{\,(N_c^2+1)-2\,\big\}\,\Phi_\partial^{\alpha}(x)
+\frac{1}{N_c^2-1}\big\{\,(N_c^2+1)-6\,\big\}\,\pi\Phi_G^\alpha(x,x)\ .
\end{align*}
The other correlators can be calculated analogously.
For $qq$ scattering we obtain:
\begin{subequations}\label{qqToddCors}
\begin{align}
\Phi_\partial^{[tt^*]\alpha}(x)
&=\Phi_\partial^{[uu^*]\alpha}(x)
=\Phi_\partial^\alpha(x)+
\frac{N_c^2-5}{N_c^2-1}\,\pi\Phi_G^\alpha(x,x)\ ,\label{qqToddCorsA}\\
%%%%
\Phi_\partial^{[tu^*]\alpha}(x)
&=\Phi_\partial^{[ut^*]\alpha}(x)
=\Phi_\partial^\alpha(x)-\frac{N_c^2+3}{N_c^2-1}\,\pi\Phi_G^\alpha(x,x)\ ,
\end{align}
\end{subequations}
and from Eq.~\eqref{QuarkAntiquark} we get for $q\bar q$ scattering
\begin{subequations}\label{qANTIqToddCors}
\begin{align}
\Phi_\partial^{[ss^*]\alpha}(x)
&=\Phi_\partial^{[st^*]\alpha}(x)
=\Phi_\partial^{[ts^*]\alpha}(x)
=\Phi_\partial^\alpha(x)+\frac{N_c^2+1}{N_c^2-1}\,\pi\Phi_G^\alpha(x,x)\ ,\\
%%%%
\Phi_\partial^{[tt^*]\alpha}(x)
&=\Phi_\partial^\alpha(x)-\frac{N_c^2-3}{N_c^2-1}\,\pi\Phi_G^\alpha(x,x)\ .
\end{align}
\end{subequations}

The integrated quark correlator $\Phi(x)$ is parametrized as follows in 
terms of quark distribution
functions~\cite{Jaffe:1991ra,Chen:1994ar,Levelt:1994np,Tangerman:1994eh}
\begin{subequations}\label{QuarkCorr}
\begin{alignat}{1}
\Phi_U(x;P)
&=\tfrac{1}{2}\,f_1(x)\slash P\ ,\\
\Phi_L(x;P)
&=\tfrac{1}{2}S_\sL\,g_1(x)\gamma_5\slash P\ ,\\
\Phi_T(x;P)
&=\tfrac{1}{2}\,h_1(x)\gamma_5\tfrac{1}{2}[\slash S_\st,\slash P]\ ,
\end{alignat}
\end{subequations}
where
\begin{equation}
\epsilon_\st^{\mu\nu}=\frac{1}{P{\cdot}n}\,\epsilon^{Pn\mu\nu}\ ,
\qquad\text{and}\qquad
S=S_\sL\frac{1}{M}\,P-S_\sL\frac{M}{2\,P{\cdot}n}\,n+S_\st\ ,
\end{equation}
with $S_\sL^2{+}S_\st^2=-1$.
The indices $U$, $L$ and $T$ refer to unpolarized, 
longitudinally and transversely polarized hadrons, respectively. For the 
$T$-even transverse momentum weighted correlator $\Phi_\partial(x)$ and
the $T$-odd gluonic pole $\pi\Phi_G(x,x)$ one has the parametrizations
\begin{subequations}\label{QuarkCorr1}
\begin{alignat}{4}
\big(\Phi_\partial^\alpha\big)_U(x;P)
&=0\ ,
&\big(\pi\Phi_G^\alpha\big)_U(x;P)
&=\tfrac{1}{2}M\,ih_1^{\perp(1)}(x)\tfrac{1}{2}[\slash P,\gamma^\alpha]\ ,\\
\big(\Phi_\partial^\alpha\big)_L(x;P)
&=\tfrac{1}{2}S_\sL\,M\,h_{1L}^{\perp(1)}(x)\gamma_5
\tfrac{1}{2}[\slash P,\gamma^\alpha]\ ,\qquad
&\big(\pi\Phi_G^\alpha\big)_L(x;P)
&=0\ ,\\
\big(\Phi_\partial^\alpha\big)_T(x;P)
&=\tfrac{1}{2}M\,S_\st^\alpha g_{1T}^{(1)}(x)\gamma_5\slash P\ ,
&\big(\pi\Phi_G^\alpha\big)_T(x;P)
&=\tfrac{1}{2}M\,\epsilon_\st^{\alpha S_\st}f_{1T}^{\perp(1)}(x)\slash P\ .
\end{alignat}
\end{subequations}
From the parametrizations given above and using the decomposition in Eq.~\eqref{qqToddCorsA},
we find that the $T$-odd distribution functions $f_{1T}^{\perp(1)}$ and $h_1^{\perp(1)}$ appear with a multiplicative prefactor
$C_G^{[tt^*]}{=}(N_c^2{-}5)/(N_c^2{-}1)$ in the contribution corresponding to the $tt^*$-channel in $qq$-scattering.
This is the appropriate generalization of the factors 
$C_G^{[\mathcal U^{[\pm]}]}=\pm1$ occurring in SIDIS and Drell-Yan 
scattering (as explained in section~\ref{CrossSec}).
Similarly, the prefactors of the $T$-odd distribution functions appearing in the other scattering channels can be read of from Eq.~\eqref{qqToddCors} for $qq$ scattering and from Eq.~\eqref{qANTIqToddCors} for $q\bar q$ scattering.
These prefactors are summarized in Table~\ref{TABLE}.
From the Eqs.~\eqref{qqToddCors} and~\eqref{qANTIqToddCors} we also see that all the $T$-even distribution functions occur in hadron-hadron scattering in the same way as they do in SIDIS,
i.e.\ with a prefactor $+1$. 
For antiquark distribution functions, which can be related to
quark distributions in the negative $x$ region,
the same results as above apply.
The antiquark distribution functions will be distinguished from their quark counterparts by an overline,
e.g.\ $\bar f_1(x)$, etc.

\begin{table}
\begin{tabular}{r|c|c|c|c|}
$\mathcal U$&$\mathcal U^{[\pm]}$&
$\mathcal U^{[\Box]}\mathcal U^{[+]}$&
$\frac{1}{N_c}\tr(\mathcal U^{[\Box]})\mathcal U^{[+]}$&
$\frac{1}{N_c}\tr(\mathcal U^{[\Box]\dagger})\mathcal U^{[+]}$\\[2pt]
\hline
$\rule{0pt}{4mm}\Phi^{[\mathcal U]}$&$\Phi^{[\pm]}$&
$\Phi^{[\Box+]}$&$\Phi^{[(\Box)+]}$&
$\Phi^{[(\Box^\dagger)+]}$\\
\hline
$C_G^{[\mathcal U]}$&
$\pm1$&$3$&$1$&$1$\\[1pt]
\hline
\end{tabular}\\[4mm]
\begin{tabular}{r|c|c|c|c|c|c|c|c|}
$\mathcal U$&
$\mathcal U_{qq}^{[tt^*]}$&$\mathcal U_{qq}^{[uu^*]}$&
$\mathcal U_{qq}^{[tu^*]}$&$\mathcal U_{qq}^{[ut^*]}$&
$\mathcal U_{q\overline q}^{[tt^*]}$&
$\mathcal U_{q\overline q}^{[ss^*]}$&
$\mathcal U_{q\overline q}^{[st^*]}$&
$\mathcal U_{q\overline q}^{[ts^*]}$\\[2pt]
\hline
$\Phi^{[\mathcal U]}$&
$\Phi^{[tt^*]}$&$\Phi^{[uu^*]}$&
$\Phi^{[tu^*]}$&$\Phi^{[ut^*]}$&
$\Phi^{[tt^*]}$&$\Phi^{[ss^*]}$&
$\Phi^{[st^*]}$&$\Phi^{[ts^*]}$\\
\hline
$\rule{0pt}{12pt}C_G^{[\mathcal U]}$&
\multicolumn{2}{|c|}{$\frac{N_c^2{-}5}{N_c^2{-}1}$}&
\multicolumn{2}{|c|}{${-}\frac{N_c^2{+}3}{N_c^2{-}1}$}&
${-}\frac{N_c^2{-}3}{N_c^2{-}1}$&
\multicolumn{3}{|c|}{$\frac{N_c^2{+}1}{N_c^2{-}1}$}\\[4pt]
\hline
\end{tabular}
\parbox{0.85\textwidth}{
\caption{
The basic gauge-links (upper table) and the gauge-links in specific 
hard scattering ($qq$ and $q\bar q$) diagrams (lower table), the
notations used for the correlators and the strengths $C_G$ of the 
gluonic pole contribution $\pi\Phi_G$.
\label{TABLE}}}
\end{table}

\section{
consequences of gauge-links for fragmentation functions\label{CONSEQUENCES2}}

The discussion on the consequences of the gauge-links for fragmentation 
functions is a little bit more involved than for distribution functions,
due to the presence of the hadronic states $\vert K, X\rangle$ in the
definition of the correlators. These are {\em out}-states, preventing
the use of time-reversal symmetry to constrain the parametrization.

All collinear interactions between the soft and hard parts result in
the quark-fragmentation correlator $\Delta^{[-]}(k)$ in SIDIS
and the correlator $\Delta^{[+]}(k)$ in electron-positron annihilation
(see equation~\eqref{QuarkFrag}).
The transverse-momentum integrated fragmentation correlators in these 
two processes are
\begin{equation}
\Delta^{[\pm]}(z)
=\int\frac{d(\xi{\cdot}K)}{2\pi}\ e^{i(\xi\cdot K)/z}
\langle0|\,U_{[\pm\infty,\xi]}^-\,\psi(\xi)\,
a_h^\dagger a_h^{\phantom{\dagger}}\,
\overline\psi(0)\,U_{[0,\pm\infty]}^-\,|0\rangle\,\big\rfloor_{\text{LC}}\ .
\end{equation}
Although not immediately evident, it is not hard to see that, 
since there are only gauge-links along the $n_h$-direction, 
the two correlators are
identical: $\Delta^{[+]}(z)=\Delta^{[-]}(z)\equiv\Delta(z)$.

In analogy to the previous appendix we define a correlator $\Delta_D^\alpha$ and a gluonic-pole matrix element $\Delta_G^\alpha$
\begin{gather}
\Delta_D^\alpha(z)
=\int\frac{d(\xi{\cdot}K)}{2\pi}\ e^{i(\xi\cdot K)/z}\,
\langle0|\,U_{[\zeta,\xi]}^-\,iD^\alpha(\xi)\,\psi(\xi)\,
a_h^\dagger a_h^{\phantom{\dagger}}\,
\overline\psi(0)\,U_{[0,\zeta]}^-\,|0\rangle\,\big\rfloor_{\text{LC}}\ ,
\displaybreak[0]\\
\Delta_G^\alpha(\tfrac{1}{z},\tfrac{1}{z}{-}\tfrac{1}{z'})
=\int\frac{d(\xi{\cdot}K)}{2\pi}\frac{d(\eta{\cdot}K)}{2\pi}\ 
e^{i(\xi\cdot K)/z}e^{i[(\eta\cdot K){-}(\xi\cdot K)]/{z'}}\nonumber\\
\mspace{300mu}\times
\langle0|\,U_{[\zeta,\eta]}^-\,gG^{n_h\alpha}(\eta)\,
U_{[\eta,\xi]}^-\,\psi(\xi)\,a_h^\dagger a_h^{\phantom{\dagger}}\,
\overline\psi(0)\,U_{[0,\zeta]}^-\,|0\rangle\,\big\rfloor_{\text{LC}}\ ,
\end{gather}
with $G^{n_h\alpha}=g_{\mu\nu}G^{\mu\alpha}n_h^\nu$ 
(and $\zeta$ an arbitrary point).
It can be shown that in terms of these the weighted correlators can be 
written as
\begin{subequations}
\begin{gather}
\Delta_\partial^{[\pm]\alpha}(z)
=\Delta_D^\alpha(z)
-\int d(\tfrac{1}{z'})\ %dz'{}^{-1}\ 
\frac{i}{\frac{1}{z'}\mp i\epsilon}\,
\Delta_G^\alpha(\tfrac{1}{z},\tfrac{1}{z}{-}\tfrac{1}{z'})
=\Delta_\partial^\alpha(z)\pm\pi\Delta_G^\alpha(\tfrac{1}{z},\tfrac{1}{z})\ ,\label{D5a}\\
\Delta_\partial^{[-\Box^\dagger]\alpha}(z)
=\Delta_D^\alpha(z)
-\int d(\tfrac{1}{z'})\ %dz'{}^{-1}\ 
\bigg\{\,\frac{i}{\frac{1}{z'}+i\epsilon}
+2\pi\delta\big(\tfrac{1}{z'}\big)\,\bigg\}\,
\Delta_G^\alpha(\tfrac{1}{z},\tfrac{1}{z}{-}\tfrac{1}{z'})
=\Delta_\partial^\alpha(z)-3\pi\Delta_G^\alpha(\tfrac{1}{z},\tfrac{1}{z})\ ,\\
\Delta_\partial^{[-(\Box)]\alpha}(z)
=\Delta_\partial^{[-(\Box^\dagger)]\alpha}(z)
=\Delta_D^\alpha(z)
-\int d(\tfrac{1}{z'})\ %dz'{}^{-1}\ 
\frac{i}{\frac{1}{z'}+i\epsilon}\,
\Delta_G^\alpha(\tfrac{1}{z},\tfrac{1}{z}{-}\tfrac{1}{z'})
=\Delta_\partial^\alpha(z)-\pi\Delta_G^\alpha(\tfrac{1}{z},\tfrac{1}{z})\ ,
\end{gather}
\end{subequations}
where $\Delta_\partial$ without link index refers to
\begin{equation}
\Delta_\partial^\alpha(z)
=\Delta_D^\alpha(z)
-\int dz'{}^{-1}\ 
P\frac{i}{z'{}^{-1}}\ 
\Delta_G^\alpha(\tfrac{1}{z},\tfrac{1}{z}{-}\tfrac{1}{z'})\ .
\end{equation}

As stated at the end of appendix~\ref{a}, 
the gauge-links in the fragmentation correlators in 
$p^\uparrow p\rightarrow\pi\pi X$ are obtained from~\eqref{QuarkQuark} and~\eqref{QuarkAntiquark} by time-reversal.
We, then, find the following quark-fragmentation correlator for the 
$tt^*$-channel in quark-quark scattering (cf.~\eqref{QuarkQuarkA}):
\begin{align*}
\Delta_\partial^{[tt^*]\alpha}(z)
&=\frac{1}{N_c^2-1}\Big\{\,
(N_c^2+1)\,\Delta_\partial^{[-(\Box^\dagger)]\alpha}(z)
-2\,\Delta_\partial^{[-\Box^\dagger]}(z)\,\Big\}\nonumber\\
&=\frac{1}{N_c^2-1}\big\{\,(N_c^2+1)-2\,\big\}\,\Delta_\partial^\alpha(z)
-\frac{1}{N_c^2-1}\big\{\,(N_c^2+1)-6\,\big\}\,\pi\Delta_G^\alpha(\tfrac{1}{z},\tfrac{1}{z})\ .
\end{align*}
The other quark-fragmentation correlators can be calculated analogously.
For quark-quark scattering we obtain:
\begin{subequations}\label{QQfragcorr}
\begin{align}
\Delta_\partial^{[tt^*]\alpha}(z)
&=\Delta_\partial^{[uu^*]\alpha}(z)
=\Delta_\partial^\alpha(z)
-\frac{N_c^2-5}{N_c^2-1}\,\pi\Delta_G^\alpha(\tfrac{1}{z},\tfrac{1}{z})\ ,
\label{QQfragcorrA}\\
\Delta_\partial^{[tu^*]\alpha}(z)
&=\Delta_\partial^{[ut^*]\alpha}(z)
=\Delta_\partial^\alpha(z)
+\frac{N_c^2+3}{N_c^2-1}\,\pi\Delta_G^\alpha(\tfrac{1}{z},\tfrac{1}{z})\ ,
\end{align}
\end{subequations}
and for quark-antiquark scattering
\begin{subequations}\label{QantiQfragcorr}
\begin{align}
\Delta_\partial^{[ss^*]\alpha}(z)
&=\Delta_\partial^{[st^*]\alpha}(z)=\Delta_\partial^{[ts^*]\alpha}(z)
=\Delta_\partial^\alpha(z)
-\frac{N_c^2+1}{N_c^2-1}\,\pi\Delta_G^\alpha(\tfrac{1}{z},\tfrac{1}{z})\ ,\\
\Delta_\partial^{[tt^*]\alpha}(z)
&=\Delta_\partial^\alpha(z)
+\frac{N_c^2-3}{N_c^2-1}\,\pi\Delta_G^\alpha(\tfrac{1}{z},\tfrac{1}{z})\ .
\end{align}
\end{subequations}

The integrated fragmentation correlator $\Delta(z)$ is parametrized 
as follows~\cite{Mulders:1995dh}
\begin{subequations}\label{FragCorr}
\begin{alignat}{1}
z\,\Delta_U(z;K)
&=D_1(z)\slash K\ ,\\
z\,\Delta_L(z;K)
&=S_\sL\,G_1(z)\gamma_5\slash K\ ,\\
z\,\Delta_T(z;K)
&=H_1(z)\gamma_5\tfrac{1}{2}[\slash S_\st,\slash K]\ ,
\end{alignat}
\end{subequations}
with 
\begin{equation}
\epsilon_\st^{\mu\nu}=\frac{1}{K{\cdot}n_h}\,\epsilon^{Kn_h\mu\nu}\ ,
\qquad\text{and}\qquad
S
=S_\sL\frac{1}{M_h}\,K-S_\sL\frac{M_h}{2\,K{\cdot}n_h}\,n_h+S_\st\ .
\end{equation}
The functions in these parametrizations are called quark fragmentation functions.
Due to the internal soft interactions in the final-state hadron the 
correlators $\Delta_\partial$ and $\pi\Delta_G$ both contain $T$-even 
\emph{and} $T$-odd parts~\cite{Boer:2003cm}.
Correspondingly, they have very similar parametrizations in terms of fragmentation functions.
We will distinguish the fragmentation functions in these two correlators by adding a tilde to the fragmentation functions appearing in the parametrization of the gluonic pole.
That is, parametrizing the correlator $\Delta_\partial$ as follows
\begin{subequations}\label{FragCorr1a}
\begin{align}
z\,\big(\Delta_\partial^\alpha\big)_U(z;K)
&=M_h\,iH_1^{\perp(1)}(z)\tfrac{1}{2}[\slash K,\gamma^\alpha]\ ,\\
z\,\big(\Delta_\partial^\alpha\big)_L(z;K)
&=S_\sL\,M_h\,H_{1L}^{\perp(1)}(z)\gamma_5
\tfrac{1}{2}[\slash K,\gamma^\alpha]\ ,\\
z\,\big(\Delta_\partial^\alpha\big)_T(z;K)
&=M_h\,\Big\{S_\st^\alpha G_{1T}^{(1)}(z)\gamma_5\slash K
-\epsilon_\st^{\alpha S_\st}D_{1T}^{\perp(1)}(z)\slash K\Big\}\ ,
\end{align}
\end{subequations}
the parametrization of the gluonic pole is written as
\begin{subequations}\label{FragCorr1b}
\begin{align}
z\,\big(\pi\Delta_G^\alpha\big)_U(\tfrac{1}{z},\tfrac{1}{z};K)
&=M_h\,i\widetilde H_1^{\perp(1)}(z)\tfrac{1}{2}[\slash K,\gamma^\alpha]\ ,\\
z\,\big(\pi\Delta_G^\alpha\big)_L(\tfrac{1}{z},\tfrac{1}{z};K)
&=S_\sL\,M_h\,\widetilde H_{1L}^{\perp(1)}(z)\gamma_5
\tfrac{1}{2}[\slash K,\gamma^\alpha]\ ,\\
z\,\big(\pi\Delta_G^\alpha\big)_T(\tfrac{1}{z},\tfrac{1}{z};K)
&=M_h\,\Big\{S_\st^\alpha \widetilde G_{1T}^{(1)}(z)\gamma_5\slash K
-\epsilon_\st^{\alpha S_\st}\widetilde D_{1T}^{\perp(1)}(z)\slash K\Big\}\ .
\end{align}
\end{subequations}
The fragmentation functions appearing in these parametrizations contribute to azimuthal asymmetries in special combinations.
For instance, using the decomposition in Eq.~\eqref{QQfragcorrA} we find that the Collins effect contributed by the $tt^*$-channel for $qq$ scattering is 
$H_1^{\perp(1)}{-}\frac{N_c^2{-}5}{N_c^2{-}1}\,\widetilde H_1^{\perp(1)}$.
Similarly, the other partonic channels contribute particular combinations of fragmentation functions.
Which combination of fragmentation functions one should take for a certain process can be read of directly from the decompositions in Eq.~\eqref{QQfragcorr} and~\eqref{QantiQfragcorr}.
That is, if we let $\text{FF}(z)$ denote a generic fragmentation function appearing in the parametrizations in Eq.~\eqref{FragCorr1a} and~\eqref{FragCorr1b},
then this fragmentation function will appear in the expressions for azimuthal asymmetries in the combination 
$\text{FF}(z){-}C_G^{[\mathcal U]}\,\widetilde{\text{FF}}(z)$.
In particular, we see that the tilde fragmentation functions always appear with the (process dependent) prefactors $C_G^{[\mathcal U]}$ summarized in
Table~\ref{TABLE2},
while the fragmentation functions without a tilde always occur with a simple prefactor $+1$.
If the gluonic pole matrix elements $\pi\Delta_G$ vanish,
then so do all the tilde functions.
In that case fragmentation is completely described by the universal functions appearing in the parametrization of $\Delta_\partial$.
Notably, the Collins effect is always given by the term $H_1^{\perp(1)}(z)$.

For antiquark-fragmentation functions, which can be related to the
quark-fragmentation functions in the negative $z$ region,
the same results as above apply.

\begin{table}
\begin{tabular}{r|c|c|c|c|}
$\mathcal U$&$\mathcal U^{[\pm]}$&
$\mathcal U^{[-]}\mathcal U^{[\Box]\dagger}$&
$\frac{1}{N_c}\mathcal U^{[-]}\tr(\mathcal U^{[\Box]})$&
$\frac{1}{N_c}\mathcal U^{[-]}\tr(\mathcal U^{[\Box]\dagger})$\\[2pt]
\hline
$\rule{0pt}{4mm}\Delta^{[\mathcal U]}$&$\Delta^{[\pm]}$&
$\Delta^{[-\Box^\dagger]}$&$\Delta^{[-(\Box)]}$&
$\Delta^{[-(\Box^\dagger)]}$\\
\hline
$C_G^{[\mathcal U]}$&
$\pm1$&$-3$&$-1$&$-1$\\[1pt]
\hline
\end{tabular}\\[4mm]
\begin{tabular}{r|c|c|c|c|c|c|c|c|}
$\mathcal U$&
$\mathcal U_{qq}^{[tt^*]}$&$\mathcal U_{qq}^{[uu^*]}$&
$\mathcal U_{qq}^{[tu^*]}$&$\mathcal U_{qq}^{[ut^*]}$&
$\mathcal U_{q\overline q}^{[tt^*]}$&
$\mathcal U_{q\overline q}^{[ss^*]}$&
$\mathcal U_{q\overline q}^{[st^*]}$&
$\mathcal U_{q\overline q}^{[ts^*]}$\\[2pt]
\hline
$\Delta^{[\mathcal U]}$&
$\Delta^{[tt^*]}$&$\Delta^{[uu^*]}$&
$\Delta^{[tu^*]}$&$\Delta^{[ut^*]}$&
$\Delta^{[tt^*]}$&$\Delta^{[ss^*]}$&
$\Delta^{[st^*]}$&$\Delta^{[ts^*]}$\\
\hline
$C_G^{[\mathcal U]}$&
$-\frac{1}{2}$&$-\frac{1}{2}$&
$\frac{3}{2}$&$\frac{3}{2}$&
$\frac{3}{4}$&$-\frac{5}{4}$&
$-\frac{5}{4}$&$-\frac{5}{4}$\\[2pt]
\hline
\end{tabular}
\parbox{0.85\textwidth}{
\caption{
The basic gauge-links (upper table) and the gauge-links in specific 
hard scattering ($qq$ and $q\bar q$) diagrams (lower table), the
notations used for the correlators and the strengths $C_G$ of the 
gluonic pole contribution $\pi\Delta_G$ with $N_c{=}3$.
\label{TABLE2}}}
\end{table}

\section{\label{diagrammaticresults}
Results in the diagrammatic approach}

In the expressions given below $y$ is given by Eq.~\eqref{Y} and $\hat s$ is
$\hat s=x_\perp^2\,s\,\cosh^2[\frac{1}{2}(\eta_1{-}\eta_2)]$
= $x_\perp^2\,s/4y(1{-}y)$.
The summations run over all quark flavors, 
including the case that $q'=q$ (where applicable).
Similarly, the $\delta^{j_iq}$ are delta functions in flavor space,
indicating that the jet $j_i$ is produced by quark $q$.
We have written the factors $C_G^{[\mathcal U]}$
for the $T$-odd distribution functions between braces $\{\,\cdot\,\}$.
For the Collins functions we have written the combinations 
$H_1^{\perp(1)}{-}C_G^{[\mathcal U]}\,\widetilde H_1^{\perp(1)}$ between braces.
The factors $C_G^{[\mathcal U]}$ are taken from Table~\ref{TABLE} for the distribution functions and from Table~\ref{TABLE2} for the fragmentation functions.

\subsection*{$p^\uparrow{+}p\rightarrow\pi{+}\pi{+}X$}

Considering only the (anti)quark contributions in 
$p^\uparrow{+}p\rightarrow\pi{+}\pi{+}X$ the resulting cross
section is given by
\begin{subequations}\label{EXPR}
\begin{gather}
\langle\,\tfrac{1}{2}\sin(\delta\phi)\,d\sigma[h_1h_2]\,\rangle\nonumber\\
\mspace{50mu}=dx_{1\perp}\,dx_{2\perp}\,d\eta_1\,d\eta_2\,
\frac{d\phi_1}{2\pi}\cos(\phi_1{-}\phi_\sS)
\int\frac{dx_\perp}{x_\perp}\,\frac{\hat s}{2x_\perp\sqrt s}\nonumber\\
\mspace{100mu}\times\,\frac{4\pi\,\alpha_\sS^2}{9\,\hat s^2}\,\bigg\{\ 
M_1\,\frac{(1-y)^2+1}{y^2}
\sum_{q,q'}\{\tfrac{1}{2}\}\,{f^q}_{1T}^{\perp(1)}(x_1)f_1^{q'}(x_2)
D_1^q(z_1)D_1^{q'}(z_2)\label{EXPRa}\\
\mspace{200mu}
-M_1\,\frac{1}{3}\frac{1}{y(1-y)}
\sum_q\{-\tfrac{3}{2}\}\,{f^q}_{1T}^{\perp(1)}(x_1)f_1^q(x_2)
D_1^q(z_1)D_1^q(z_2)\label{EXPRb}\displaybreak[0]\\
\mspace{200mu}
+M_1\,\big((1-y)^2+y^2\big)
\sum_{q,q'}\{\tfrac{5}{4}\}\,{f^q}_{1T}^{\perp(1)}(x_1)\bar f{}_1^q(x_2)
D_1^{q'}(z_1)\bar D{}_1^{q'}(z_2)\label{EXPRc}\displaybreak[0]\\
\mspace{200mu}
+M_1\,\frac{(1-y)^2+1}{y^2}
\sum_{q,q'}\{-\tfrac{3}{4}\}\,{f^q}_{1T}^{\perp(1)}(x_1)
\bar f{}_1^{q'}(x_2)
D_1^{q}(z_1)\bar D{}_1^{q'}(z_2)\label{EXPRd}\displaybreak[0]\\
\mspace{200mu}
+M_1\,\frac{2}{3}\frac{(1-y)^2}{y}
\sum_q\{\tfrac{5}{4}\}\,{f^q}_{1T}^{\perp(1)}(x_1)
\bar f{}_1^q(x_2)
D_1^{q}(z_1)\bar D{}_1^q(z_2)\label{EXPRe}\displaybreak[0]\\
\mspace{200mu}
-M_2\,2y(1-y)
\sum_{q,q'}\{\tfrac{5}{4}\}\,h_1^q(x_1){\bar h{}^q}_1^{\perp(1)}(x_2)
D_1^{q'}(z_1)\bar D{}_1^{q'}(z_2)\label{EXPRf}\displaybreak[0]\\
\mspace{200mu}
-M_2\,\frac{2}{3}(1-y)
\sum_q\{\tfrac{5}{4}\}\,h_1^q(x_1){\bar h{}^q}_1^{\perp(1)}(x_2)
D_1^{q}(z_1)\bar D{}_1^q(z_2)\label{EXPRg}\displaybreak[0]\\
\mspace{200mu}
-M_2\,\frac{1}{3}
\sum_q\{-\tfrac{3}{2}\}\,h_1^q(x_1){h^q}_1^{\perp(1)}(x_2)
D_1^q(z_1)D_1^q(z_2)\label{EXPRh}\displaybreak[0]\\
\mspace{200mu}
-M_{h_1}\,2\frac{1-y}{y^2}
\sum_{q,q'}h_1^q(x_1)f_1^{q'}(x_2)\,
\big\{H^q{}_1^{\perp(1)}(z_1)
{-}\tfrac{1}{2}\widetilde H^q{}_1^{\perp(1)}(z_1)\,\big\}\,
D_1^{q'}(z_2)\label{EXPRi}\displaybreak[0]\\
\mspace{200mu}
+M_{h_1}\,\frac{2}{3}\frac{1}{y}
\sum_qh_1^q(x_1)f_1^q(x_2)\,
\big\{H^q{}_1^{\perp(1)}(z_1)
{+}\tfrac{3}{2}\widetilde H^q{}_1^{\perp(1)}(z_1)\,\big\}\,
D_1^q(z_2)\label{EXPRj}\displaybreak[0]\\
\mspace{200mu}
-M_{h_1}\,2\frac{1-y}{y^2}
\sum_{q,q'}h_1^q(x_1)\bar f{}_1^{q'}(x_2)\,
\big\{H^q{}_1^{\perp(1)}(z_1)
{+}\tfrac{3}{4}\widetilde H^q{}_1^{\perp(1)}(z_1)\,\big\}\,
\bar D{}_1^{q'}(z_2)\label{EXPRk}\displaybreak[0]\\
\mspace{200mu}
-M_{h_1}\,\frac{2}{3}\frac{1-y}{y}
\sum_qh_1^q(x_1)\bar f{}_1^q(x_2)\,
\big\{H^q{}_1^{\perp(1)}(z_1)
{-}\tfrac{5}{4}\widetilde H^q{}_1^{\perp(1)}(z_1)\,\big\}\,
\bar D{}_1^q(z_2)\label{EXPRl}\\
\mspace{200mu}
+M_{h_1}\frac{2}{3}
\sum_qh_1^q(x_1)\bar f{}_1^q(x_2)\,
\big\{\bar H^q{}_1^{\perp(1)}(z_1)
{-}\tfrac{5}{4}\widetilde{\bar H}{}^q{}_1^{\perp(1)}(z_1)\,\big\}\,
D_1^q(z_2)\label{EXPRm}\\
\mspace{300mu}
+\big(\,\text{quarks}\leftrightarrow\text{antiquarks}\,\big)\ \ 
\bigg\}\ %\nonumber\\
%%
%\mspace{300mu}
+\big(\,K_1\leftrightarrow K_2\,\big)\nonumber%\\
\end{gather}
\end{subequations}

\subsection*{$p+p\rightarrow\pi+\text{Jet}+X$}

Considering only the (anti)quark contributions in 
$p^\uparrow{+}p\rightarrow\pi{+}\text{Jet}{+}X$ the resulting cross
section is given by
\begin{subequations}\label{EXPR2}
\begin{gather}
\langle\,\tfrac{1}{2}\sin(\delta\phi)\,d\sigma[h_1j_2]\,\rangle\nonumber\\
\mspace{50mu}=dx_{1\perp}\,dx_{2\perp}\,d\eta_1\,d\eta_2\,
\frac{d\phi_1}{2\pi}\cos(\phi_1{-}\phi_\sS)\,
\frac{\hat s}{2x_{2\perp}\sqrt s}\,\nonumber\\
\mspace{100mu}\times\,\frac{4\pi\,\alpha_\sS^2}{9\,\hat s^2}\,\bigg\{\ 
M_1\,\frac{(1-y)^2+1}{y^2}
\sum_{q,q'}\{\tfrac{1}{2}\}\,{f^q}_{1T}^{\perp(1)}(x_1)f_1^{q'}(x_2)
D_1^q(z_1)\,\delta^{j_2q'}\label{EXPR2a}\\
\mspace{200mu}
+M_1\,\frac{y^2+1}{(1-y)^2}
\sum_{q,q'}\{\tfrac{1}{2}\}\,{f^q}_{1T}^{\perp(1)}(x_1)f_1^{q'}(x_2)
D_1^{q'}(z_1)\,\delta^{j_2q}\label{EXPR2b}\\
\mspace{200mu}
-M_1\,\frac{2}{3}\frac{1}{y(1-y)}
\sum_q\{-\tfrac{3}{2}\}\,{f^q}_{1T}^{\perp(1)}(x_1)f_1^q(x_2)
D_1^q(z_1)\,\delta^{j_2q}\label{EXPR2c}\displaybreak[0]\\
\mspace{200mu}
+M_1\,\big((1-y)^2+y^2\big)
\sum_{q,q'}\{\tfrac{5}{4}\}\,{f^q}_{1T}^{\perp(1)}(x_1)\bar f{}_1^q(x_2)
D_1^{q'}(z_1)\,\delta^{j_2\bar q'}\label{EXPR2d}\displaybreak[0]\\
\mspace{200mu}
+M_1\,\big((1-y)^2+y^2\big)
\sum_{q,q'}\{\tfrac{5}{4}\}\,{f^q}_{1T}^{\perp(1)}(x_1)\bar f{}_1^q(x_2)
\bar D{}_1^{q'}(z_1)\,\delta^{j_2q'}\label{EXPR2e}\displaybreak[0]\\
\mspace{200mu}
+M_1\,\frac{(1-y)^2+1}{y^2}
\sum_{q,q'}\{-\tfrac{3}{4}\}\,{f^q}_{1T}^{\perp(1)}(x_1)
\bar f{}_1^{q'}(x_2)
D_1^{q}(z_1)\,\delta^{j_2\bar q'}\label{EXPR2f}\displaybreak[0]\\
\mspace{200mu}
+M_1\,\frac{y^2+1}{(1-y)^2}
\sum_{q,q'}\{-\tfrac{3}{4}\}\,{f^q}_{1T}^{\perp(1)}(x_1)
\bar f{}_1^{q'}(x_2)
\bar D{}_1^{q'}(z_1)\,\delta^{j_2q}\label{EXPR2g}\displaybreak[0]\\
\mspace{200mu}
+M_1\,\frac{2}{3}\frac{(1-y)^2}{y}
\sum_q\{\tfrac{5}{4}\}\,{f^q}_{1T}^{\perp(1)}(x_1)
\bar f{}_1^q(x_2)
D_1^{q}(z_1)\,\delta^{j_2\bar q}\label{EXPR2h}\displaybreak[0]\\
\mspace{200mu}
+M_1\frac{2}{3}\frac{y^2}{1-y}
\sum_q\{\tfrac{5}{4}\}\,{f^q}_{1T}^{\perp(1)}(x_1)\bar f{}_1^q(x_2)
\bar D{}_1^q(z_1)\,\delta^{j_2q}\label{EXPR2i}\displaybreak[0]\\
\mspace{200mu}
-M_2\,2y(1-y)
\sum_{q,q'}\{\tfrac{5}{4}\}\,h_1^q(x_1)\bar h{}^q{}_1^{\perp(1)}(x_2)
D_1^{q'}(z_1)\,\delta^{j_2\bar q'}\label{EXPR2j}\displaybreak[0]\\
\mspace{200mu}
-M_2\,2y(1-y)
\sum_{q,q'}\{\tfrac{5}{4}\}\,h_1^q(x_1)\bar h{}^q{}_1^{\perp(1)}(x_2)
\bar D{}_1^{q'}(z_1)\,\delta^{j_2q'}\label{EXPR2k}\displaybreak[0]\\
\mspace{200mu}
-M_2\,\frac{2}{3}(1-y)
\sum_q\{\tfrac{5}{4}\}\,h_1^q(x_1){\bar h{}^q}_1^{\perp(1)}(x_2)
D_1^{q}(z_1)\,\delta^{j_2\bar q}\label{EXPR2l}\displaybreak[0]\\
\mspace{200mu}
-M_2\frac{2}{3}y
\sum_q\{\tfrac{5}{4}\}\,h_1^q(x_1)\bar h{}^q{}_1^{\perp(1)}(x_2)
\bar D{}_1^q(z_1)\,\delta^{j_2q}\label{EXPR2m}\displaybreak[0]\\
\mspace{200mu}
-M_2\,\frac{2}{3}
\sum_q\{-\tfrac{3}{2}\}\,h_1^q(x_1){h^q}_1^{\perp(1)}(x_2)
D_1^q(z_1)\,\delta^{j_2q}\label{EXPR2n}\displaybreak[0]\\
\mspace{200mu}
-M_{h_1}\,2\frac{1-y}{y^2}
\sum_{q,q'}h_1^q(x_1)f_1^{q'}(x_2)\,
\big\{H^q{}_1^{\perp(1)}(z_1)
{-}\tfrac{1}{2}\widetilde H^q{}_1^{\perp(1)}(z_1)\,\big\}\,\delta^{j_2q'}
\label{EXPR2o}\displaybreak[0]\\
\mspace{200mu}
+M_{h_1}\,\frac{2}{3}\frac{1}{y}
\sum_qh_1^q(x_1)f_1^q(x_2)\,
\big\{H^q{}_1^{\perp(1)}(z_1)
{+}\tfrac{3}{2}\widetilde H^q{}_1^{\perp(1)}(z_1)\,\big\}\,\delta^{j_2q}
\label{EXPR2p}\displaybreak[0]\\
\mspace{200mu}
-M_{h_1}\,2\frac{1-y}{y^2}
\sum_{q,q'}h_1^q(x_1)\bar f{}_1^{q'}(x_2)\,
\big\{H^q{}_1^{\perp(1)}(z_1)
{+}\tfrac{3}{4}\widetilde H^q{}_1^{\perp(1)}(z_1)\,\big\}\,\delta^{j_2\bar q'}
\label{EXPR2q}\displaybreak[0]\\
\mspace{200mu}
-M_{h_1}\,\frac{2}{3}\frac{1-y}{y}
\sum_qh_1^q(x_1)\bar f{}_1^q(x_2)\,
\big\{H^q{}_1^{\perp(1)}(z_1)
{-}\tfrac{5}{4}\widetilde H^q{}_1^{\perp(1)}(z_1)\,\big\}\,\delta^{j_2\bar q}
\label{EXPR2r}\\
\mspace{200mu}
+M_{h_1}\frac{2}{3}
\sum_qh_1^q(x_1)\bar f{}_1^q(x_2)\,
\big\{\bar H^q{}_1^{\perp(1)}(z_1)
{-}\tfrac{5}{4}\widetilde{\bar H}{}^q{}_1^{\perp(1)}(z_1)\,\big\}\,\delta^{j_2q}
\label{EXPR2s}\\
\mspace{300mu}
+\big(\,\text{quarks}\leftrightarrow\text{antiquarks}\,\big)\ \ 
\bigg\}\nonumber
\end{gather}
\end{subequations}

\subsection*{$p+p\rightarrow\text{Jet}+\text{Jet}+X$}

Considering only the (anti)quark contributions in 
$p^\uparrow{+}p\rightarrow\text{Jet}{+}\text{Jet}{+}X$ the resulting cross
section is given by
\begin{subequations}\label{EXPR3}
\begin{gather}
\langle\,\tfrac{1}{2}\sin(\delta\phi)\,d\sigma[j_1j_2]\,\rangle\nonumber\\
\mspace{50mu}=dx_{1\perp}\,dx_{2\perp}\,d\eta_1\,d\eta_2\,
\frac{d\phi_1}{2\pi}\cos(\phi_1{-}\phi_\sS)\,
\delta(x_{1\perp}{-}x_{2\perp})\,
\frac{\hat s}{2\sqrt s}\,\nonumber\\
\mspace{100mu}\times\,\frac{4\pi\,\alpha_\sS^2}{9\,\hat s^2}\,\bigg\{\ 
M_1\,\frac{(1-y)^2+1}{y^2}
\sum_{q,q'}\{\tfrac{1}{2}\}\,{f^q}_{1T}^{\perp(1)}(x_1)f_1^{q'}(x_2)
\,\delta^{j_1q}\delta^{j_2q'}\label{EXPR3a}\\
\mspace{200mu}
-M_1\,\frac{1}{3}\frac{1}{y(1-y)}
\sum_q\{-\tfrac{3}{2}\}\,{f^q}_{1T}^{\perp(1)}(x_1)f_1^q(x_2)
\,\delta^{j_1q}\delta^{j_2q}\label{EXPR3c}\displaybreak[0]\\
\mspace{200mu}
+M_1\,\big((1-y)^2+y^2\big)
\sum_{q,q'}\{\tfrac{5}{4}\}\,{f^q}_{1T}^{\perp(1)}(x_1)\bar f{}_1^q(x_2)
\,\delta^{j_1q'}\delta^{j_2\bar q'}\label{EXPR3d}\displaybreak[0]\\
\mspace{200mu}
+M_1\,\frac{(1-y)^2+1}{y^2}
\sum_{q,q'}\{-\tfrac{3}{4}\}\,{f^q}_{1T}^{\perp(1)}(x_1)
\bar f{}_1^{q'}(x_2)
\,\delta^{j_1q}\delta^{j_2\bar q'}\label{EXPR3f}\displaybreak[0]\\
\mspace{200mu}
+M_1\,\frac{2}{3}\frac{(1-y)^2}{y}
\sum_q\{\tfrac{5}{4}\}\,{f^q}_{1T}^{\perp(1)}(x_1)
\bar f{}_1^q(x_2)
\,\delta^{j_1q}\delta^{j_2\bar q}\label{EXPR3h}\displaybreak[0]\\
\mspace{200mu}
-M_2\,2y(1-y)
\sum_{q,q'}\{\tfrac{5}{4}\}\,h_1^q(x_1)\bar h{}^q{}_1^{\perp(1)}(x_2)
\,\delta^{j_1q'}\delta^{j_2\bar q'}\label{EXPR3j}\displaybreak[0]\\
\mspace{200mu}
-M_2\,\frac{2}{3}(1-y)
\sum_q\{\tfrac{5}{4}\}\,h_1^q(x_1){\bar h{}^q}_1^{\perp(1)}(x_2)
\,\delta^{j_1q}\delta^{j_2\bar q}\label{EXPR3l}\displaybreak[0]\\
\mspace{200mu}
-M_2\,\frac{1}{3}
\sum_q\{-\tfrac{3}{2}\}\,h_1^q(x_1){h^q}_1^{\perp(1)}(x_2)
\,\delta^{j_1q}\delta^{j_2q}\label{EXPR3n}\\
\mspace{300mu}
+\big(\,\text{quarks}\leftrightarrow\text{antiquarks}\,\big)\ \ \bigg\}\ 
+\big(\,\text{Jet}_1\leftrightarrow\text{Jet}_2\,\big)\nonumber%\\
\end{gather}
\end{subequations}

\section{Partonic cross sections\label{PARTONX}}

In this appendix we enumerate all the (anti)quark scattering cross sections 
(taken from~\cite{Bacchetta:2004it}) that are needed in this paper.

\subsection*{Quark-quark scattering}

The unpolarized quark-quark scattering cross sections are given by
\begin{subequations}
\begin{gather}
\frac{d\hat \sigma_{qq^\prime\rightarrow qq^\prime}}{d\hat t} 
=\frac{4\pi\alpha_\sS^2}{9\,\hat s^2}\,
\frac{\hat s^2 + \hat u^2}{\hat t^2}\ ,\\
%%%%
\frac{d\hat \sigma_{qq^\prime\rightarrow q^\prime q}}{d\hat t} 
=\frac{4\pi\alpha_\sS^2}{9\,\hat s^2}\,
\frac{\hat s^2 + \hat t^2}{\hat u^2}\ ,\\
%%%%
\frac{d\hat \sigma_{qq\rightarrow qq}}{d\hat t} 
=\frac{d\hat \sigma_{qq^\prime\rightarrow qq^\prime}}{d\hat t}
+\frac{d\hat \sigma_{qq^\prime\rightarrow q^\prime q}}{d\hat t}
-2\,\frac{d\hat \sigma^{\rm I}_{qq\rightarrow qq}}{d\hat t}\ ,
\end{gather}
\end{subequations}
where $d\hat\sigma^{\rm I}$ represents the interference terms
\begin{equation}
\frac{d\hat \sigma^{\rm I}_{qq\rightarrow qq}}{d\hat t}
=\frac{4\pi\alpha_\sS^2}{27\,\hat s^2}\,\frac{\hat s^2}{\hat t\hat u}\ .
\end{equation}
The polarized quark-quark scattering cross sections are
\begin{subequations}
\begin{gather}
\frac{d\Delta\hat\sigma_{q^\uparrow q^\uparrow\rightarrow qq}}{d\hat t}
=-\frac{8\pi\alpha_\sS^2}{27\,\hat s^2}\ ,\\
%%%%
\frac{d\Delta\hat\sigma_{q^\uparrow q'\rightarrow q^\uparrow q'}}{d\hat t}
=-\frac{8\pi\alpha_\sS^2}{9\,\hat s^2}\,\frac{\hat u\hat s}{\hat t^2}\ ,\\
%%%%
\frac{d\Delta\hat\sigma_{q^\uparrow q\rightarrow q^\uparrow q}}{d\hat t}
=\frac{d\Delta\hat\sigma_{q^\uparrow q'\rightarrow q^\uparrow q'}}{d\hat t}
-\frac{d\Delta\hat\sigma_{q^\uparrow q\rightarrow q^\uparrow q}^{\rm I}}
{d\hat t}\ ,
\end{gather}
\end{subequations}
with the interference term
\begin{equation}
%%%%
\frac{d\Delta\hat\sigma_{q^\uparrow q\rightarrow q^\uparrow q}^{\rm I}}
{d\hat t}
=-\frac{8\pi\alpha_\sS^2}{27\,\hat s^2}\,\frac{\hat s}{\hat t}\ .
\end{equation}
The modified cross sections are
\begin{subequations}
\begin{gather}
\frac{d\hat\sigma_{\widehat{gq}q'\rightarrow qq'}}{d\hat t}
=\frac{N_c^2{-}5}{N_c^2{-}1}\,
\frac{d\hat\sigma_{qq'\rightarrow qq'}}{d\hat t}\ ,\\
%%%%
\frac{d\hat\sigma_{\widehat{gq}q'\rightarrow q'q}}{d\hat t}
=\frac{N_c^2{-}5}{N_c^2{-}1}\,
\frac{d\hat\sigma_{qq'\rightarrow q'q}}{d\hat t}\ ,\displaybreak[0]\\
%%%%
\frac{d\hat\sigma_{\widehat{gq}q\rightarrow qq}}{d\hat t} 
=\frac{N_c^2{-}5}{N_c^2{-}1}\,
\bigg[\,\frac{d\hat \sigma_{qq^\prime\rightarrow qq^\prime}}{d\hat t}
+\frac{d\hat \sigma_{qq^\prime\rightarrow q^\prime q}}{d\hat t}\,\bigg] 
+2\,\frac{N_c^2{+}3}{N_c^2{-}1}\,
\frac{d\hat \sigma^{\rm I}_{qq\rightarrow qq}}{d\hat t}\ ,\linebreak[0]\\
%%%%
\frac{d\Delta\hat\sigma_{q^\uparrow\widehat{gq}^\uparrow{\rightarrow}qq}}
{d\hat t}
=-\frac{N_c^2{+}3}{N_c^2{-}1}\,
\frac{d\Delta\hat\sigma_{q^\uparrow q^\uparrow\rightarrow qq}}
{d\hat t}\ ,\displaybreak[0]\\
%%%%
\frac{d\Delta\hat\sigma_{q^\uparrow q'
\rightarrow\widehat{gq}{}^\uparrow q'}}{d\hat t}
=-\frac{N_c^2{-}5}{N_c^2{-}1}\,
\frac{d\Delta\hat\sigma_{q^\uparrow q'\rightarrow q^\uparrow q'}}
{d\hat t}\ ,\\
%%%%
\frac{d\Delta\hat\sigma_{q^\uparrow q
\rightarrow\widehat{gq}{}^\uparrow q}}{d\hat t}
=-\frac{N_c^2{-}5}{N_c^2{-}1}\,
\frac{d\Delta\hat\sigma_{q^\uparrow q'\rightarrow q^\uparrow q'}}{d\hat t}
-\frac{N_c^2{+}3}{N_c^2{-}1}\,
\frac{d\Delta\hat\sigma_{q^\uparrow q\rightarrow q^\uparrow q}^{\mathrm I}}
{d\hat t}\ .
%\\
%%%%%
%\frac{d\hat\sigma_{qq\rightarrow\widehat{gq}q}}{d\hat t} 
%=-\frac{N_c^2{-}5}{N_c^2{-}1}\,
%\bigg[\,\frac{d\hat \sigma_{qq^\prime\rightarrow qq^\prime}}{d\hat t}
%+\frac{d\hat \sigma_{qq^\prime\rightarrow q^\prime q}}{d\hat t}\,\bigg]
%-2\,\frac{N_c^2{+}3}{N_c^2{-}1}\,
%\frac{d\hat \sigma^{\rm I}_{qq\rightarrow qq}}{d\hat t}\ .
\end{gather}
\end{subequations}
The partonic cross sections can be regarded as functions of the variable 
$y$ defined in~\eqref{Y} through
\begin{equation}
\frac{\hat t}{\hat s}=-y\ ,\qquad
\frac{\hat u}{\hat s}=-(1-y)\ ,\qquad
\hat s=\frac{x_\perp^2\,s}{4y(1-y)}\ .
\end{equation}
All other non-vanishing quark-quark and antiquark-antiquark scattering cross sections that contribute to hadronic pion production can be obtained from the above from symmetry considerations.

\subsection*{Quark-antiquark scattering}

The unpolarized quark-antiquark scattering cross sections are given by
\begin{subequations}
\begin{gather}
\frac{d\hat \sigma_{q\bar q^\prime\rightarrow q\bar q^\prime}}{d\hat t} 
=\frac{4\pi\alpha_\sS^2}{9\,\hat s^2}\,
\frac{\hat s^2 + \hat u^2}{\hat t^2}\ ,\\
%%%%
\frac{d\hat \sigma_{q\bar q \rightarrow q^\prime \bar q^\prime}}{d\hat t} 
=\frac{4\pi\alpha_\sS^2}{9\,\hat s^2}\,\frac{\hat t^2 + \hat u^2}{\hat s^2}\ ,\\
%%%%
\frac{d\hat \sigma_{q\bar q\rightarrow q\bar q}}{d\hat t} 
=\frac{d\hat \sigma_{q\bar q^\prime\rightarrow q\bar q^\prime}}{d\hat t}
+\frac{d\hat \sigma_{q\bar q \rightarrow q^\prime \bar q^\prime}}{d\hat t}
-2\,\frac{d\hat \sigma^{\rm I}_{q\bar q\rightarrow q\bar q}}{d\hat t}\ ,
\end{gather}
\end{subequations}
with the interference term
\begin{equation}
\frac{d\hat \sigma^{\rm I}_{q\bar q\rightarrow q\bar q}}{d\hat t}
\equiv\frac{4\pi\alpha_\sS^2}{27\,\hat s^2}\,\frac{\hat u^2}{\hat t\hat s}\ .
\end{equation}
The polarized quark-antiquark scattering cross sections are
\begin{subequations}
\begin{gather}
\frac{d\Delta\hat\sigma_{q^\uparrow\bar q^\uparrow\rightarrow q'\bar q'}}
{d\hat t}
=-\frac{8\pi\alpha_\sS^2}{9\,\hat s^2}\,\frac{\hat t\hat u}{\hat s^2}\ ,\\
%%%%
\frac{d\Delta\hat\sigma_{q^\uparrow\bar q^\uparrow\rightarrow q\bar q}}{d\hat t}
=\frac{d\Delta\hat\sigma_{q^\uparrow\bar q^\uparrow\rightarrow q'\bar q'}}
{d\hat t}
-\frac{d\Delta\hat\sigma_{q^\uparrow\bar q^\uparrow\rightarrow q\bar q}^{\rm I}}
{d\hat t}\ ,\displaybreak[0]\\
%%%%
\frac{d\Delta\hat\sigma_{q^\uparrow\bar q'\rightarrow q^\uparrow\bar q'}}
{d\hat t}
=-\frac{8\pi\alpha_\sS^2}{9\,\hat s^2}\,\frac{\hat u\hat s}
{\hat t^2}\ ,\displaybreak[0]\\
%%%%
\frac{d\Delta\hat\sigma_{q^\uparrow\bar q\rightarrow q^\uparrow\bar q}}{d\hat t}
=\frac{d\Delta\hat\sigma_{q^\uparrow\bar q'\rightarrow q^\uparrow\bar q'}}
{d\hat t}
-\frac{d\Delta\hat\sigma_{q^\uparrow\bar q\rightarrow q^\uparrow\bar q}^{\rm I}}
{d\hat t}\ ,\\
%%%%
\frac{d\Delta\hat\sigma_{q^\uparrow\bar q\rightarrow\bar q^\uparrow q}}{d\hat t}
=-\frac{8\pi\alpha_\sS^2}{27\hat s^2}\ ,
\end{gather}
\end{subequations}
with the interference terms
\begin{equation}
\frac{d\Delta\hat\sigma_{q^\uparrow\bar q^\uparrow\rightarrow q\bar q}^{\rm I}}
{d\hat t}
=-\frac{8\pi\alpha_\sS^2}{27\,\hat s^2}\,\frac{\hat u}{\hat s}\ ,\qquad
%%%%
\frac{d\Delta\hat\sigma_{q^\uparrow\bar q\rightarrow q^\uparrow\bar q}^{\rm I}}
{d\hat t}
=-\frac{8\pi\alpha_\sS^2}{27\hat s^2}\,\frac{\hat u}{\hat t}\ .
\end{equation}
The modified cross sections are
\begin{subequations}
\begin{gather}
\frac{d\hat\sigma_{\widehat{gq}\bar q'\rightarrow q\bar q'}}{d\hat t}
=-\frac{N_c^2{-}3}{N_c^2{-}1}\,
\frac{d\hat\sigma_{q\bar q'\rightarrow q\bar q'}}{d\hat t}\ ,\\
%%%%
\frac{d\hat\sigma_{\widehat{gq}\bar q\rightarrow q'\bar q'}}{d\hat t}
=\frac{N_c^2{+}1}{N_c^2{-}1}\,
\frac{d\hat\sigma_{q\bar q\rightarrow q'\bar q'}}{d\hat t}\ ,\displaybreak[0]\\
%%%%
\frac{d\hat\sigma_{\widehat{gq}\bar q\rightarrow q\bar q}}{d\hat t} 
=-\frac{N_c^2{-}3}{N_c^2{-}1}\,
\frac{d\hat \sigma_{q\bar q^\prime\rightarrow q\bar q^\prime}}{d\hat t}
+\frac{N_c^2{+}1}{N_c^2{-}1}\bigg[\,
\frac{d\hat \sigma_{q\bar q \rightarrow q^\prime \bar q^\prime}}{d\hat t}
-2\,\frac{d\hat \sigma^{\rm I}_{q\bar q\rightarrow q\bar q}}{d\hat t}\,
\bigg]\ ,\linebreak[0]\\
%%%%
\frac{d\Delta\hat\sigma_{q^\uparrow\widehat{g\bar q}{}^\uparrow
\rightarrow q'\bar q'}}{d\hat t}
=\frac{N_c^2{+}1}{N_c^2{-}1}\,
\frac{d\Delta\hat\sigma_{q^\uparrow\bar q{}^\uparrow\rightarrow q'\bar q'}}
{d\hat t}\ ,\displaybreak[0]\\
%%%%
\frac{d\Delta\hat\sigma_{q^\uparrow\widehat{g\bar q}{}^\uparrow
\rightarrow q\bar q}}{d\hat t}
=\frac{N_c^2{+}1}{N_c^2{-}1}\,
\frac{d\Delta\hat\sigma_{q^\uparrow\bar q{}^\uparrow\rightarrow q\bar q}}
{d\hat t}\ ,\displaybreak[0]\\
%%%%
\frac{d\Delta\hat\sigma_{q^\uparrow\bar q'
\rightarrow\widehat{gq}{}^\uparrow\bar q'}}{d\hat t}
=\frac{N_c^2{-}3}{N_c^2{-}1}\,
\frac{d\Delta\hat\sigma_{q^\uparrow\bar q'\rightarrow q^\uparrow\bar q'}}
{d\hat t}\ ,\displaybreak[0]\\
%%%%
\frac{d\Delta\hat\sigma_{q^\uparrow\bar q
\rightarrow\widehat{gq}{}^\uparrow\bar q}}{d\hat t}
=\frac{N_c^2{-}3}{N_c^2{-}1}\,
\frac{d\Delta\hat\sigma_{q^\uparrow\bar q'\rightarrow q^\uparrow\bar q'}}
{d\hat t}
+\frac{N_c^2{+}1}{N_c^2-1}\,
\frac{d\Delta\hat\sigma_{q^\uparrow\bar q
\rightarrow q^\uparrow\bar q}^{\mathrm I}}{d\hat t}\ ,\\
%%%%
\frac{d\Delta\hat\sigma_{q^\uparrow\bar q
\rightarrow\widehat{g\bar q}{}^\uparrow q}}{d\hat t}
=-\frac{N_c^2{+}1}{N_c^2{-}1}\,
\frac{d\Delta\hat\sigma_{q^\uparrow\bar q\rightarrow\bar q{}^\uparrow q}}
{d\hat t}\ .
%\\
%%%%%
%\frac{d\Delta\hat\sigma_{q\bar q\rightarrow \widehat{gq}\bar q}}{d\hat t} 
%= \frac{N_c^2{-}3}{N_c^2{-}1}\,
%\frac{d\hat \sigma_{q\bar q^\prime\rightarrow q\bar q^\prime}}{d\hat t}
%-\frac{N_c^2{+}1}{N_c^2{-}1}\bigg[\,
%\frac{d\hat \sigma_{q\bar q \rightarrow q^\prime \bar q^\prime}}{d\hat t}
%-2\,\frac{d\hat \sigma^{\rm I}_{q\bar q\rightarrow q\bar q}}{d\hat t}\,
%\bigg]\ .
\end{gather}
\end{subequations}
All other non-vanishing quark-antiquark scattering cross sections that contribute to hadronic pion production can be obtained from the above from symmetry considerations.

\begin{acknowledgments}
We acknowledge discussions with D.~Boer and W.~Vogelsang. 
Part of this work was supported by the foundation for Fundamental 
Research of Matter (FOM) and the National Organization for Scientific 
Research (NWO). The work of A.B. was supported by the A. von Humboldt Foundation.
\end{acknowledgments}

\bibliographystyle{apsrev}
\bibliography{references}

\end{document}